\documentclass[reprint,amsmath,amssymb,aps,prb,superscriptaddress]{revtex4-2}
\usepackage{array}
\usepackage{bm}
\usepackage{color}
\usepackage{float}
\usepackage{graphicx}
\usepackage{hyperref}
\hypersetup{colorlinks=true,allcolors=blue}
\usepackage{natbib}
\usepackage{newtxtext}
\usepackage{newtxmath}
\usepackage[caption=false]{subfig}

\begin{document}

\title{A pseudofermion functional renormalization group study of dipolar-octupolar pyrochlore magnets}

\author{Li Ern Chern}
\affiliation{T.C.M. Group, Cavendish Laboratory, University of Cambridge, Cambridge CB3 0HE, United Kingdom}

\author{F\'{e}lix Desrochers}
\affiliation{Department of Physics, University of Toronto, Toronto, Ontario M5S 1A7, Canada}

\author{Yong Baek Kim}
\affiliation{Department of Physics, University of Toronto, Toronto, Ontario M5S 1A7, Canada}

\author{Claudio Castelnovo}
\affiliation{T.C.M. Group, Cavendish Laboratory, University of Cambridge, Cambridge CB3 0HE, United Kingdom}


\begin{abstract}
Motivated by recent experiments on Ce$_2$Zr$_2$O$_7$ that reveal a dynamic, liquid-like ground state, we study the nearest neighbor XYZ Hamiltonian of dipolar-octupolar pyrochlore magnets with the pseudofermion functional renormalization group (PFFRG), which is numerically implemented by the \textit{SpinParser} software. Taking the interaction between the octupolar components to be dominant and antiferromagnetic, we map out the phase diagram demarcating the quantum disordered and magnetically ordered states. We identify four distinct phases, namely the $0$-flux and $\pi$-flux quantum spin ices, and the all-in-all-out magnetic orders along the local $z$ and $x$ axes. We further use the static two-spin correlations output by the PFFRG algorithm to compute the polarized neutron scattering cross-sections, which are able to capture several qualitative features observed experimentally, in the materially relevant parameter regime that stabilizes the $\pi$-flux quantum spin ice. Our results provide support for a quantum spin liquid ground state in Ce$_2$Zr$_2$O$_7$.
\end{abstract}


\maketitle


\section{\label{section:introduction}Introduction}

A spin liquid is, roughly speaking, a cooperative paramagnet with disordered yet correlated spins. Depending on whether its dynamics are governed by thermal or quantum fluctuations, a spin liquid is primarily classified as classical or quantum. A prominent example of spin liquids is spin ice \cite{spinicebook} on the three-dimensional pyrochlore lattice, where each unit tetrahedron displays a two-in-two-out spin configuration. While there exist concrete experimental evidences of classical spin ice \cite{science.1064761,nature06433,annurev-conmatphys-020911-125058} in Ho$_2$Ti$_2$O$_7$ and Dy$_2$Ti$_2$O$_7$ \cite{nature20619,PhysRevLett.87.047205,nphys1227,science.1178868,science.1177582}, the detection and confirmation of quantum spin ice (QSI) \cite{PhysRevB.69.064404,PhysRevLett.108.037202,PhysRevB.86.075154,Gingras_2014} in candidate materials remains an ongoing work. QSI is of fundamental theoretical interest as it realizes a lattice analog of quantum electrodynamics, which exhibits gapless photons and two types of gapped excitations, namely spinons and visons, which are the electric and magnetic monopoles of the emergent $U(1)$ gauge theory.

It is against this backdrop that recent experiments on Ce-based pyrochlore magnets, which hinted at the stabilization of a QSI, potentially of an octupolar nature, have garnered much attention. Measurements of heat capacity, magnetic susceptibility, and muon spin relaxation in Ce$_2$Zr$_2$O$_7$ find no magnetic ordering or spin freezing down to the $\sim 10\, \mathrm{mK}$ temperature range, while inelastic neutron scattering reveals a broad and diffusive continuum \cite{PhysRevLett.122.187201,s41567-019-0577-6,PhysRevX.12.021015,PhysRevB.106.094425,PhysRevB.108.054438,2308.02800}. Similar experiments on Ce$_2$Sn$_2$O$_7$ \cite{PhysRevLett.115.097202,s41567-020-0827-7,2211.15140,2304.05452} and Ce$_2$Hf$_2$O$_7$ \cite{PhysRevMaterials.6.044406,2305.08261} also point to fluid-like ground states. In these materials, the interplay of spin-orbit coupling and crystal electric field leads to a Kramers doublet ground state for each Ce$^{3+}$ ion, which is well separated from the first excited state at $\sim 50 \, \mathrm{meV}$ \cite{PhysRevLett.122.187201,s41567-019-0577-6,s41567-020-0827-7}. The low-energy description thus consists of interacting pseudospin-$1/2$ moments, where, remarkably, the $x$ and $z$ components transform as magnetic dipoles, while the $y$ component transforms as a magnetic octupole \cite{PhysRevLett.112.167203}. As a consequence, the most generic symmetry-allowed Hamiltonian at the nearest neighbor level can be reduced to an XYZ Hamiltonian with only diagonal couplings \cite{PhysRevLett.112.167203}, the relatively simple expression of which appeals to a multitude of theoretical analyses. So far, the pyrochlore XYZ model has been studied with various mean field theories \cite{PhysRevLett.112.167203,PhysRevB.95.041106,PhysRevB.96.085136,PhysRevResearch.2.013334,PhysRevB.102.104408,PhysRevB.105.035149,PhysRevB.107.064404,2301.05240,2304.05452}, quantum Monte Carlo \cite{PhysRevResearch.2.042022}, exact diagonalization \cite{PhysRevResearch.2.023253,PhysRevB.102.104408,s41535-022-00458-2,PhysRevLett.129.097202}, and numeric linked cluster \cite{PhysRevX.12.021015,PhysRevB.108.054438} in the quantum limit, as well as classical Monte Carlo \cite{s41535-022-00458-2}, molecular dynamics \cite{s41535-022-00458-2,PhysRevLett.129.097202,PhysRevX.12.021015,PhysRevB.108.054438}, self consistent Gaussian approximation \cite{s41535-022-00458-2}, and linear spin wave theory \cite{PhysRevResearch.2.013334,PhysRevB.102.245102} in the (semi)classical limit.

In this paper, we theoretically investigate the $S=1/2$ XYZ model of the dipolar-octupolar pyrochlore magnets using the pseudofermion functional renormalization group (PFFRG) \cite{PhysRevB.81.144410,PhysRevB.83.024402,PhysRevB.84.014417,PhysRevB.84.100406,PhysRevB.89.020408,PhysRevB.94.140408,PhysRevB.94.224403,PhysRevB.94.235138,PhysRevB.96.045144,PhysRevB.97.064415,PhysRevB.97.064416,PhysRevX.9.011005,PhysRevB.100.125164,PhysRevResearch.2.013370,PhysRevB.103.184407,PhysRevB.105.054426,PhysRevResearch.5.L012025,JPSJ.92.064708,2307.10359,reutherthesis,buessenthesis}, which is numerically implemented by the \textit{SpinParser} software \cite{SciPostPhysCodeb.5,SciPostPhysCodeb.5-r1.0}. As the name itself suggests, the two essential components of PFFRG are (i) a pseudofermion representation of the spin operator, so that a Hamiltonian with arbitrary two-spin interactions is cast into an interacting fermion problem, followed by (ii) a functional renormalization group \cite{WETTERICH199390,frgbook,RevModPhys.84.299} analysis. The purpose is to obtain a low-energy description of the system in terms of the renormalized self-energy and two-particle vertex function, which are used to compute the static component of the magnetic susceptibility. Compared to other methods, PFFRG has the advantage of treating the thermodynamic limit by effectively truncating the interaction range of fermions, and it is free of the sign problem encountered in quantum Monte Carlo. Therefore, PFFRG is advocated as a suitable tool to study frustrated magnetism in three spatial dimensions.

\begin{figure}
\includegraphics[scale=0.3]{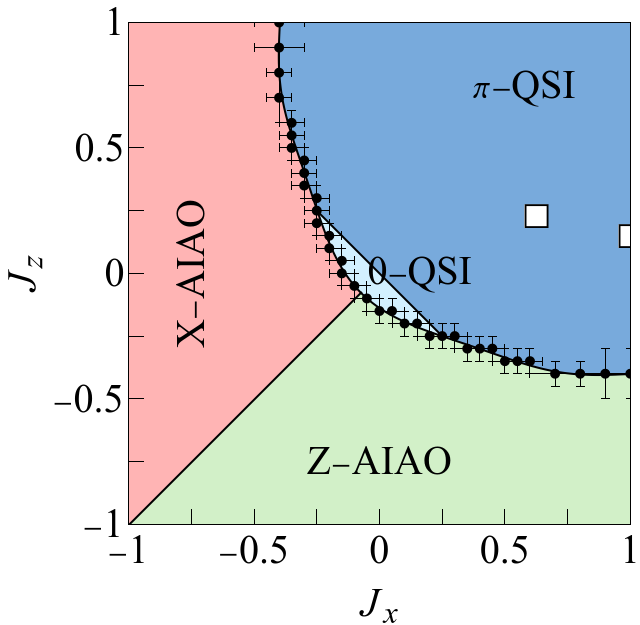}
\caption{\label{figure:phase}Phase diagram of the $S=1/2$ nearest neighbor XYZ model \eqref{xyzhamiltonian} on the pyrochlore lattice, in the parameter space where $J_y=1$ and $-1 \leq J_x, J_z \leq 1$, obtained by the pseudofermion functional renormalization group. The labels $0$-QSI, $\pi$-QSI, Z(X)-AIAO represent the $0$-flux quantum spin ice, the $\pi$-flux quantum spin ice, and the all-in-all-out magnetic order along the local $z$($x$) axes, respectively. The empty squares indicate the parametrizations of Ce$_2$Zr$_2$O$_7$ proposed by Refs.~\cite{PhysRevX.12.021015,s41535-022-00458-2} at the level of nearest neighbor interactions. The black dots demarcate the symmetry-preserving and symmetry-breaking regions as signalled by the breakdown of smooth renormalization group flow, with the error bars mainly reflecting the resolution of the grid for the calculations. The lines (without dots) separating the two spin ices and the two magnetic orders are obtained by further examining the diagonal components of the static susceptibility. See Sec.~\ref{section:phase} for more details.}
\end{figure}

We first map out a phase diagram in the parameter space relevant to Ce$_2$Zr$_2$O$_7$, as shown in Fig.~\ref{figure:phase}. Tracking the evolution of the static susceptibility with respect to the RG cutoff, PFFRG is able to distinguish symmetry-breaking ordered states and symmetry-preserving paramagnetic states \cite{PhysRevB.100.125164,SciPostPhysCodeb.5,buessenthesis}, the latter of which are putative quantum spin liquids at low temperatures. To further differentiate among the quantum spin liquids, we look for qualitative differences in the momentum-resolved static susceptibilities across the parameter space, and use insights gained from previous theoretical analyses \cite{PhysRevResearch.2.023253,PhysRevB.102.104408,2301.05240}. Within the parameter space of interest, our PFFRG analysis identifies four distinct phases: a $0$-flux QSI, a $\pi$-flux QSI, and two all-in-all-out (AIAO) magnetic orders along the local $z$ and $x$ axes, respectively \footnote{The origin in Fig.~\ref{figure:phase}, which corresponds to the classical spin ice, lies on the phase boundary between the two quantum spin ices.}. The proposed parametrizations of Ce$_2$Zr$_2$O$_7$ in the existing literature \cite{PhysRevX.12.021015,PhysRevB.108.054438,s41535-022-00458-2} are found to lie within the $\pi$-flux QSI phase.

Although the static susceptibility is not exactly the equal-time spin structure factor, the former is a good estimate of the latter if the spectral weight of the dynamical spin structure factor is concentrated at low energies, or the dynamical spin structure factor is nonzero only within a relatively narrow range of finite energies. In principle, one can calculate the magnetic susceptibility at arbitrary (Matsubara) frequency, but the integration over frequency to obtain the equal-time spin structure factor leads to additional numerical errors \cite{2307.10359}. As an approximation, we thus use the static susceptibility itself to compute the neutron scattering cross-section. We back this approximation with calculations of the static susceptibility and the equal-time spin structure factor using gauge mean field theory \cite{PhysRevB.107.064404,2301.05240}, which reveal highly similar intensity profiles. We find that the PFFRG calculated neutron scatterings at several parameters stabilizing the $\pi$-flux QSI not only reproduce the rod motifs in the spin-flip channel, but also capture the positions of the intensity minima in the non-spin-flip channel, as seen in the experimental data reported by Ref.~\cite{PhysRevX.12.021015}. Our results thus offer support to the case of a quantum spin liquid ground state in Ce$_2$Zr$_2$O$_7$, which may be the $\pi$-flux QSI.

The rest of this paper is organized as follows. Sec.~\ref{section:model} describes the XYZ model of dipolar-octupolar pyrochlore magnets. Sec.~\ref{section:method} briefly explains the PFFRG methodology, in particular how to distinguish ordered and disordered states by analyzing the cutoff dependence of the static susceptibility. Sec.~\ref{section:result} presents the results from the PFFRG analysis, namely the different phases and the calculated structure factors. Sec.~\ref{section:discussion} summarizes our work, discusses our results in light of other existing works, and comments on possible future directions.

\section{\label{section:model}Model}

Dipolar-octupolar pyrochlore magnets such as Ce$_2$Zr$_2$O$_7$ consist of interacting pseudospin-$1/2$ degrees of freedom, of which the $y$ ($x$ and $z$) components transform as magnetic octupoles (dipoles) under lattice symmetries and time reversal \cite{PhysRevLett.112.167203}, on a three-dimensional network of corner-sharing tetrahedra. The most generic nearest neighbor Hamiltonian is given by \cite{PhysRevLett.112.167203}
\begin{equation} \label{generichamiltonian}
H = \sum_{\langle ij \rangle} [ J_x S_i^x S_j^x + J_y S_i^y S_j^y + J_z S_i^z S_j^z + J_{xz} ( S_i^x S_j^z + S_i^z S_j^x ) ],
\end{equation}
where the pseudospin-$1/2$ components are defined according to the local coordinates at each site, in which the $z$ axis points from the center of a tetrahedron to one of its vertices. The off-diagonal term $J_{xz}$ in \eqref{generichamiltonian} can be removed by a rotation about the $y$ axis, which results in the XYZ model,
\begin{equation} \label{xyzhamiltonian}
H_\mathrm{XYZ} = \sum_{\langle ij \rangle} ( \tilde{J}_x \tilde{S}_i^x \tilde{S}_j^x + \tilde{J}_y \tilde{S}_i^y \tilde{S}_j^y + \tilde{J}_z \tilde{S}_i^z \tilde{S}_j^z ) .
\end{equation}
We will drop the tildes in \eqref{xyzhamiltonian} as it is suggested that $J_{xz} \approx 0$ in the candidate material \cite{PhysRevX.12.021015,s41535-022-00458-2,PhysRevB.108.054438}. Previous analyses on Ce$_2$Zr$_2$O$_7$ propose that $J_y$ is dominant and antiferromagnetic, including the possibility of $J_x \approx J_y$ \cite{PhysRevX.12.021015,s41535-022-00458-2,PhysRevB.108.054438}. Here we shall set $J_y = 1$ as the unit for energy, and study the XYZ model in the parameter regime where $\lvert J_x \rvert , \lvert J_z \rvert \leq J_y$. Further considerations about the general parameter regime $J_x,J_y,J_z \geq 0$ are given in Appendix \ref{appendix:width}.

The pyrochlore lattice is a face-centered cubic (fcc) lattice with four sites per unit cell. We use the following convention for the primitive translation vectors
\begin{equation} \label{primitivetranslate}
\mathbf{a}_1 = \frac{a}{2} (\hat{\mathbf{y}} + \hat{\mathbf{z}}), \mathbf{a}_2 = \frac{a}{2} (\hat{\mathbf{z}} + \hat{\mathbf{x}}), \mathbf{a}_3 = \frac{a}{2} (\hat{\mathbf{x}} + \hat{\mathbf{y}}) ,
\end{equation}
and the sublattice displacements
\begin{equation} \label{sublatticedisplace}
\begin{aligned}[b]
& \mathbf{d}_0 = \frac{a}{8}(+\hat{\mathbf{x}}+\hat{\mathbf{y}}+\hat{\mathbf{z}}) , \mathbf{d}_1 = \frac{a}{8}(+\hat{\mathbf{x}}-\hat{\mathbf{y}}-\hat{\mathbf{z}}) , \\
& \mathbf{d}_2 = \frac{a}{8}(-\hat{\mathbf{x}}+\hat{\mathbf{y}}-\hat{\mathbf{z}}) ,
\mathbf{d}_3 = \frac{a}{8}(-\hat{\mathbf{x}}-\hat{\mathbf{y}}+\hat{\mathbf{z}}) ,
\end{aligned}
\end{equation}
where $\hat{\mathbf{x}}, \hat{\mathbf{y}}, \hat{\mathbf{z}}$ are three orthonormal vectors in the global frame.

\section{\label{section:method}Method}

We employ the pseudofermion functional renormalization group (PFFRG) \cite{PhysRevB.81.144410,PhysRevB.83.024402,PhysRevB.84.014417,PhysRevB.84.100406,PhysRevB.89.020408,PhysRevB.94.140408,PhysRevB.94.224403,PhysRevB.94.235138,PhysRevB.96.045144,PhysRevB.97.064415,PhysRevB.97.064416,PhysRevX.9.011005,PhysRevB.100.125164,PhysRevResearch.2.013370,PhysRevB.103.184407,PhysRevB.105.054426,PhysRevResearch.5.L012025,JPSJ.92.064708,2307.10359,reutherthesis,buessenthesis} to study the $S=1/2$ XYZ Hamiltonian \eqref{xyzhamiltonian} on the pyrochlore lattice. We only provide a brief description of the method in this section, with some further details given in Appendix \ref{appendix:pffrg}. Interested readers may refer to, e.g., Refs.~\cite{PhysRevB.81.144410,PhysRevB.100.125164,2307.10359,SciPostPhysCodeb.5,reutherthesis,buessenthesis}, for an in-depth discussion.

\subsection{\label{section:pffrg}Pseudofermion Functional Renormalization Group}

We start by representing the spins in terms of pseudofermions,
\begin{equation} \label{pseudofermionrepresentation}
S_i^\mu = \frac{1}{2} \sum_{\alpha \beta} f_{i \alpha}^\dagger \sigma_{\alpha \beta}^\mu f_{i \beta} ,
\end{equation}
where $\sigma^\mu$ are Pauli matrices. A Hamiltonian with arbitrary two-spin interactions can be cast into the following form,
\begin{equation} \label{pseudofermionhamiltonian}
\begin{aligned}[b]
H &= \sum_{ij} \sum_{\mu \nu} J_{ij}^{\mu \nu} S_i^\mu S_j^\nu \\
&= \sum_{ij} \sum_{\mu \nu} \sum_{\alpha \beta \gamma \delta} \frac{J_{ij}^{\mu \nu}}{4} \sigma_{\alpha \beta}^\mu \sigma_{\gamma \delta}^\nu f_{i \alpha}^\dagger f_{j \gamma}^\dagger f_{j \delta} f_{i \beta} .
\end{aligned}
\end{equation}
Eq.~\eqref{pseudofermionrepresentation} is a faithful representation only when the single occupancy constraint $f_{i \uparrow}^\dagger f_{i \uparrow} + f_{i \downarrow}^\dagger f_{i \downarrow} = 1$  is satisfied locally. To a good approximation, this constraint is enforced on average by setting the chemical potential to be zero, as its violation leads to energetically unfavorable $S=0$ local defects, which are thermally suppressed at low temperatures \cite{PhysRevB.103.184407,SciPostPhysCodeb.5,buessenthesis}.

The strongly interacting pseudofermion Hamiltonian \eqref{pseudofermionhamiltonian} is then subject to the functional renormalization group (FRG) analysis \cite{WETTERICH199390,frgbook,RevModPhys.84.299}. The central objects of FRG are one-line irreducible vertex functions, or simply \textit{vertices}, which encode the effective $n$-particle interactions \cite{negeleorlandbook,2307.10359}. The FRG flow equations are generated by introducing an infrared cutoff $\Lambda$ in the Matsubara frequency to the bare propagator $G_0 (i \omega) = 1 / i \omega$, such that
\begin{equation} \label{barepropagator}
G_0^\Lambda (i \omega) = \frac{\theta (\lvert \omega \rvert - \Lambda)}{i \omega} .
\end{equation}
With the magnetic couplings of the original spin model treated as bare interactions at $\Lambda \longrightarrow \infty$, we are ultimately interested in the low-energy effective theory at $\Lambda \longrightarrow 0$. Differentiating the generating functional of the vertices with respect to the cutoff, one obtains an infinite hierarchy of coupled integro-differential equations, in which the flow of the $n$-particle vertex involves vertices up to the $(n+1)$-th order. We then apply the Katanin truncation scheme \cite{PhysRevB.70.115109}, so that we only have to solve the flow equations for the one-particle vertex, which is equal to the self-energy up to a minus sign, and the two-particle vertex while partially retaining the effect of the three-particle vertex \cite{reutherthesis,buessenthesis}. The structures of these vertices are considerably simplified by the symmetry of the original spin model as well as the gauge redundancy from the pseudofermion construction \cite{PhysRevB.100.125164,buessenthesis}. For instance, the self-energy is an imaginary and antisymmetric function that depends only on the Matsubara frequency. More details can be found in Appendix \ref{appendix:pffrg}.

\subsection{\label{section:numeric}Numerical Implementation}

All PFFRG calculations in this work are performed with the newly available \textit{SpinParser} software \cite{SciPostPhysCodeb.5,SciPostPhysCodeb.5-r1.0}, which solves the flow equations numerically. As shown in Appendix \ref{appendix:pffrg}, the flow equations involve summations over Matsubara frequencies and lattice sites, see \eqref{selfenergyflow} and \eqref{vertexflow}. The Matsubara frequency becomes continuous in the zero temperature limit, while the number of lattice sites grows to infinity in the thermodynamic limit. Further approximations are thus required for a numerical solution.

To this end, the frequency axis is rediscretized such that frequency dependent quantities are evaluated for a finite set of frequencies, supplemented with a linear interpolation scheme. We choose $N_\omega = 144$ frequencies distributed logarithmically around $\omega=0$, with $\lvert \omega \rvert_\mathrm{max} = 100$ and $\lvert \omega \rvert_\mathrm{min} = 0.001$. The numerical integration over frequency is performed with a trapezoidal scheme \cite{SciPostPhysCodeb.5}. On the other hand, we set the two-particle vertex to be zero if the two lattice sites involved are further apart than $L=6$ nearest neighbor bonds. Such a finite truncation range allows us to effectively study the thermodynamic limit without imposing specific boundary conditions. Finally, the cutoff is decreased in steps via $\Lambda_{n+1} = b \Lambda_n$ with the factor $b<1$. We choose $b = 0.98$, and the initial (final) cutoff at $\Lambda_i = 100$ ($\Lambda_f = 0.01$), which is much greater (smaller) than any intrinsic energy scale. All these specifications can be done within \textit{SpinParser} \cite{SciPostPhysCodeb.5,SciPostPhysCodeb.5-r1.0}.

\subsection{\label{section:susceptibility}Magnetic Susceptibility}

A useful physical observable that can be calculated from PFFRG is the static component ($i \omega=0$) of the magnetic susceptibility (two-spin correlator),
\begin{equation} \label{magneticsusceptibility}
\chi_{ij}^{\mu \nu} (i \omega) = \int_0^\beta \mathrm{d} \tau \, e^{i \omega \tau} \langle T_\tau S_i^\mu (\tau) S_j^\nu ( 0 ) \rangle .
\end{equation}
We emphasize that the magnetic susceptibility is also a function of the cutoff $\Lambda$, though it is not shown explicitly.

Analyzing the Fourier-transformed static susceptibility
\begin{equation} \label{fouriertransform}
\chi^{\mu \nu} (\mathbf{k}) = \frac{1}{N} \sum_{ij} \chi_{ij}^{\mu \nu} (i \omega = 0) e^{i \mathbf{k} \cdot (\mathbf{r}_i - \mathbf{r}_j)}
\end{equation}
allows us to infer the ground state of the system as follows. The PFFRG calculation assumes the full symmetry of the Hamiltonian, while a magnetically ordered state corresponds to spontaneous symmetry breaking. The onset of magnetic order causes a breakdown of the FRG flow, which typically manifests as a divergence or a kink in the cutoff dependence of the magnetic susceptibility \cite{SciPostPhysCodeb.5,2307.10359}. To determine the ground state, one can trace the evolution of $\chi (\mathbf{k}) \equiv \sum_{\mu} \chi^{\mu \mu} ( \mathbf{k} )$ as $\Lambda$ decreases at some momentum $\mathbf{k} = \mathbf{k}_*$, which is typically chosen such that $\chi (\mathbf{k}_*)$ is largest. If $\chi (\mathbf{k}_*)$ becomes nonanalytic at some critical cutoff $\Lambda_\mathrm{c}$, then it signals a transition into a symmetry broken phase, and $\mathbf{k_*}$ right before the flow breakdown is taken as the ordering wavevector. In contrast, if $\chi (\mathbf{k}_*)$ remains smooth and finite down to $\Lambda \longrightarrow 0$, then it indicates a paramagnetic ground state that preserves all symmetries. It is worth remarking that the solutions of the flow equations for $\Lambda < \Lambda_\mathrm{c}$ are no longer physically meaningful due to symmetry breaking, so one would not be able to access the true ground state at $\Lambda \longrightarrow 0$ should there be a multistep ordering process \cite{PhysRevB.103.184407}.

\begin{figure*}
\includegraphics[scale=0.36]{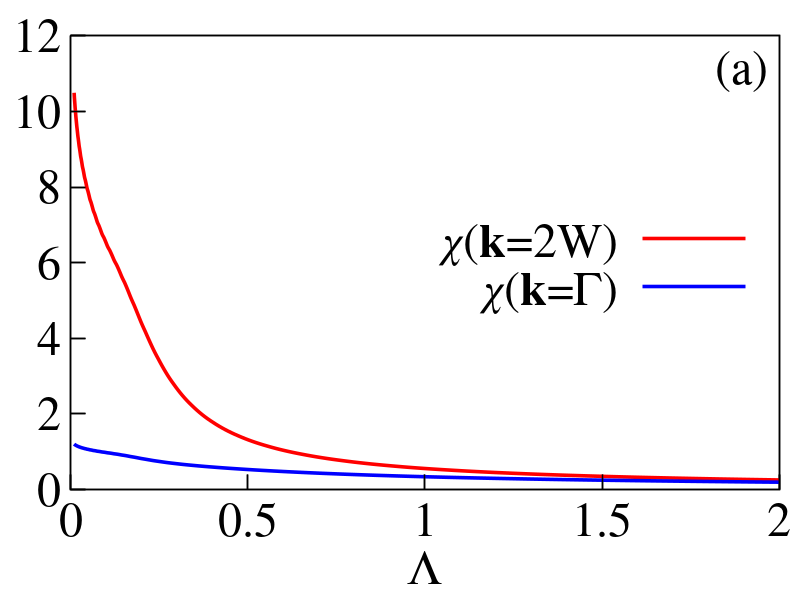} \quad
\includegraphics[scale=0.36]{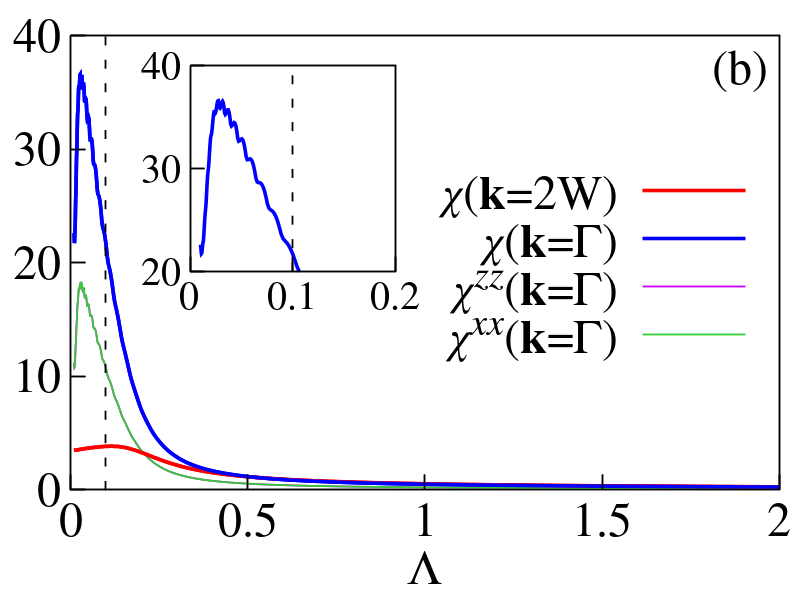} \quad
\includegraphics[scale=0.36]{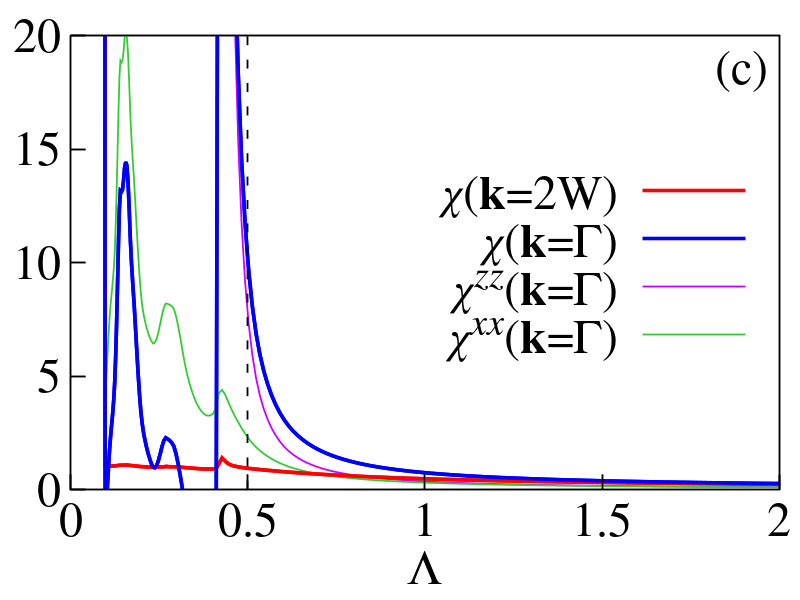} \\
\includegraphics[scale=0.36]{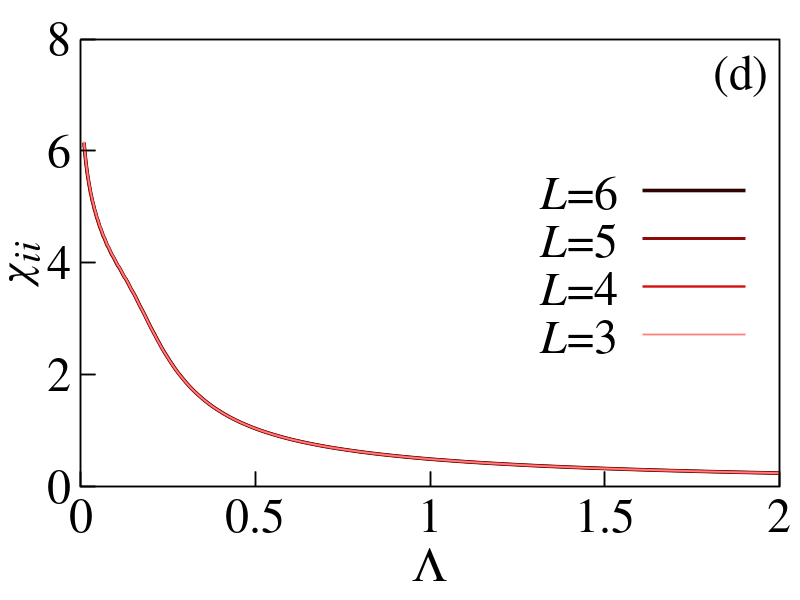} \quad
\includegraphics[scale=0.36]{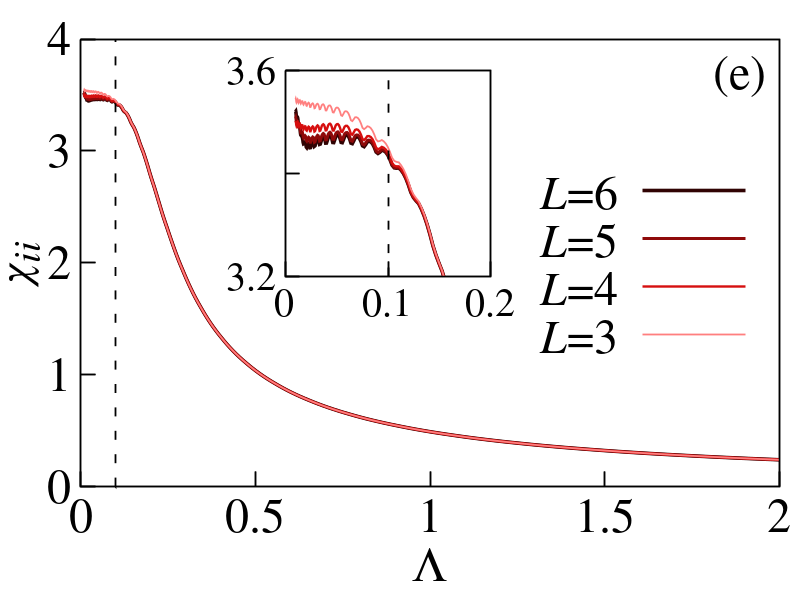} \quad
\includegraphics[scale=0.36]{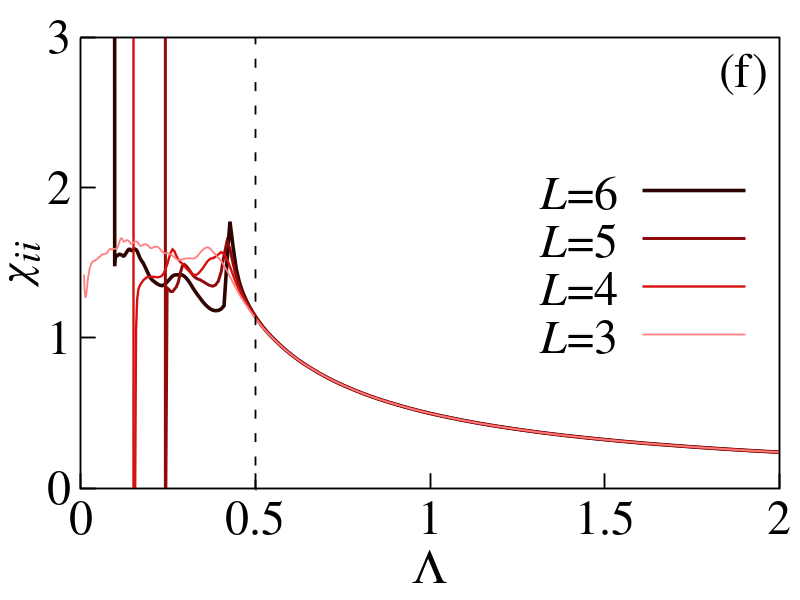}
\caption{\label{figure:evolve}Fourier-transformed static susceptibility $\chi (\mathbf{k})$ as a function of the cutoff $\Lambda$ at momenta $\mathbf{k}=\Gamma = \mathbf{0}$ and $2\mathrm{W} = (2 \pi / a, 4 \pi / a, 0)$, with the truncation range $L=6$, for $(J_x, J_z)$ equal to (a) $(0.3,0.1)$, (b) $(-0.2,-0.2)$, and (c) $(-0.5,-0.6)$. The $xx$ and $zz$ components of $\chi^{\mu \nu} (\mathbf{k} = \Gamma)$ are also plotted in (b), where they overlap, and (c). On-site susceptibility $\chi_{ii}$ as a function of the cutoff $\Lambda$ with the truncation ranges $L=3,4,5,6$, for $(J_x, J_z)$ equal to (d) $(0.3,0.1)$, (e) $(-0.2,-0.2)$, and (f) $(-0.5,-0.6)$. The data in (d) overlap with each other. Dashed lines in (b,e) and (c,f) represent estimates of the critical cutoffs $\Lambda_\mathrm{c}$ based on the discrepancy between $\chi_{ii}$ for $L=3$ and $6$, see Sec.~\ref{section:susceptibility} for details. Insets in (b,e) are zoom-ins of the data for $\Lambda \leq 0.2$.}
\end{figure*}

Some degree of uncertainty is inevitable in locating the critical cutoff $\Lambda_\mathrm{c}$ by inspection. Moreover, it may be difficult to detect the nonanalyticity in $\chi (\mathbf{k}_*)$ when the phase transition takes place at some small cutoff value $\Lambda \approx 0$. To complement the procedure described in the previous paragraph, we compare the on-site susceptibility $\chi_{ii} \equiv \sum_{\mu} \chi_{ii}^{\mu \mu} (i \omega = 0)$ as a function of $\Lambda$ for multiple truncation ranges $L$ \cite{PhysRevResearch.2.013370,PhysRevB.103.184407}. The rationale is that paramagnets and spin liquids only exhibit short-range correlations, so that $\chi_{ii}$ converges already at small $L$. In contrast, long-range correlations become important when there is a tendency for magnetic ordering, which results in an unambiguous discrepancy between $\chi_{ii}$ for small and large $L$. One can then set a quantitative threshold for the discrepancy to define $\Lambda_\mathrm{c}$ \cite{PhysRevB.103.184407}.

We illustrate these ideas with three parameters, $(J_x, J_z) = (0.3,0.1)$, $(-0.2,-0.2)$, and $(-0.5,-0.6)$. We plot the Fourier-transformed static susceptibilities $\chi (\mathbf{k})$ at momenta $\mathbf{k}=\Gamma, 2 \mathrm{W}$ \footnote{For the all-in-all-out magnetic orders, $\chi (\mathbf{k})$ peaks at $\mathbf{k} = \Gamma$. For the quantum spin ices, we find that $\chi (\mathbf{k}=2W)$ is close to, but not exactly, the maximum, which is sufficient for our analysis; the maximum seems to take place at some incommensurate wave vector that weakly depends on the couplings.} and the on-site susceptibilities $\chi_{ii}$ with the truncation ranges $L=3,4,5,6$ as functions of the cutoff $\Lambda$ in Figs.~\ref{figure:evolve}a-\ref{figure:evolve}f. At $(J_x, J_z)=(0.3,0.1)$, $\chi (\mathbf{k})$ evolves smoothly down to the smallest $\Lambda$, while $\chi_{ii}$ for different $L$ overlap almost perfectly with each other. These point to a spin liquid ground state. At $(J_x,J_z)=(-0.2,-0.2)$, $\chi (\mathbf{k}=\Gamma)$ becomes dominant as $\Lambda$ decreases, and begins to display rugged features below $\Lambda_\mathrm{c} \approx 0.1$. A small but unambiguous discrepancy in $\chi_{ii}$ between different $L$ is also seen below $\Lambda_\mathrm{c}$. These indicate a phase transition into a magnetic order. At $(J_x,J_z)=(-0.5,-0.6)$, $\chi (\mathbf{k}=\Gamma)$ grows rapidly around $\Lambda = 0.5$ and tends to diverge. In fact, one can easily see the flow instability at small $\Lambda$, where $\chi (\mathbf{k}=\Gamma)$ oscillates wildly and even goes negative. This again indicates a phase transition into a magnetic order. To estimate the critical cutoff, we calculate the relative difference $\lvert \chi_{ii} (L=6) - \chi_{ii} (L=3) \rvert / \lvert \chi_{ii} (L=6) \rvert$ and check when it reaches $1 \%$ \cite{PhysRevB.103.184407}, which gives $\Lambda_\mathrm{c} \approx 0.5$.

\section{\label{section:result}Results}

\subsection{\label{section:phase}Phase Diagram}

We have outlined in Sec.~\ref{section:susceptibility} how magnetic orders are distinguished from paramagnetic phases such as quantum spin liquids in general PFFRG calculations. We now specialize to the pyrochlore XYZ model \eqref{xyzhamiltonian}, and discuss how to further differentiate \textit{among} magnetic orders, as well as quantum spin liquids.

\begin{figure*}
\includegraphics[scale=0.23]{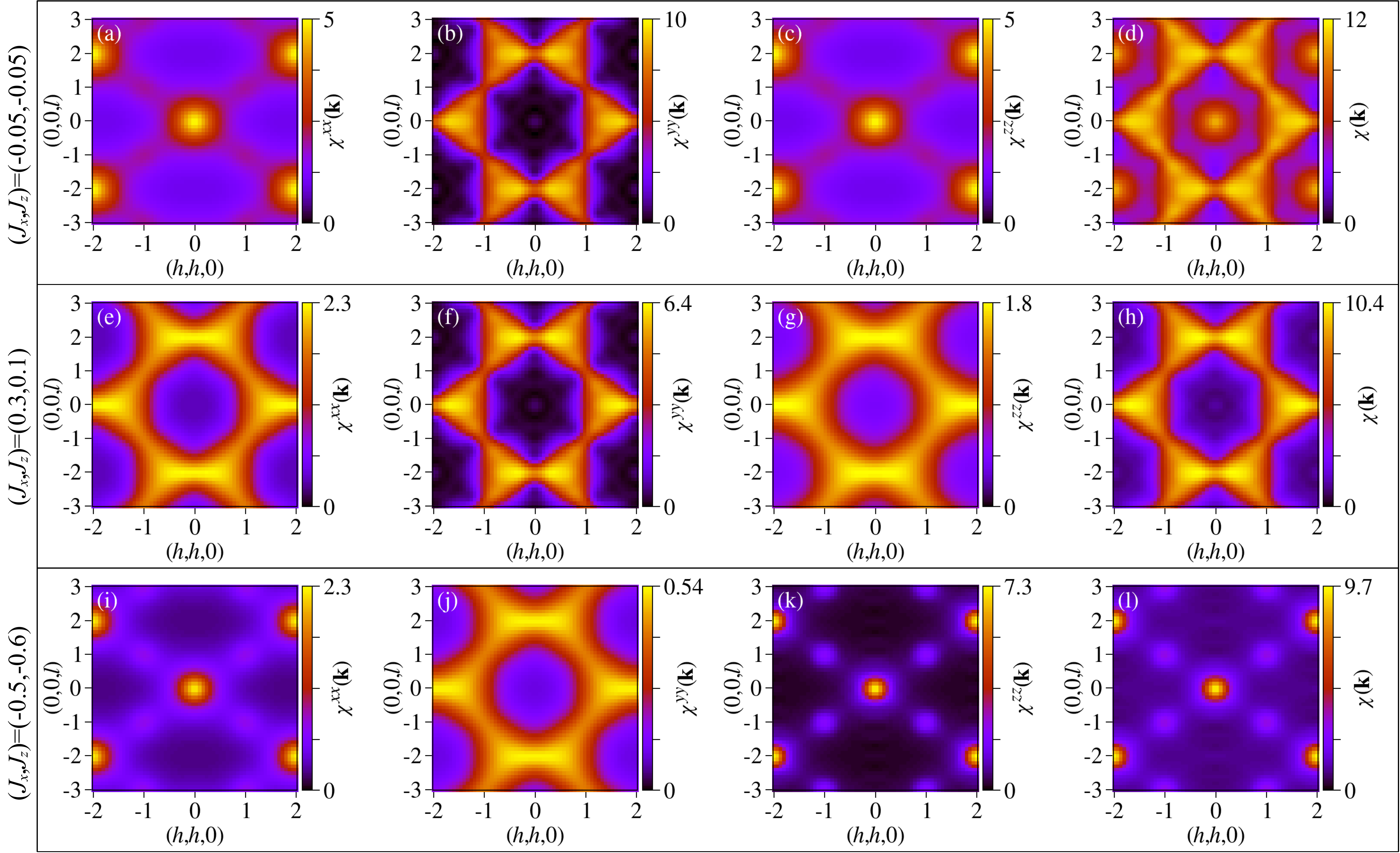}
\caption{\label{figure:correlatemain}Diagonal components of the static susceptibility, as well as their summation, in the $[hhl]$ plane, at $(J_x, J_z)$ equal to (a-d) $(-0.05, -0.05)$, (e-h) $(0.3,0.1)$, and (i-l) $(-0.5,-0.6)$, which stabilize the $0$-flux quantum spin ice, the $\pi$-flux quantum spin ice, and the all-in-all-out magnetic order, respectively.}
\end{figure*}

When a symmetry breaking phase transition occurs, $\mathbf{k} = \mathbf{k}_*$ that yields the maximum of $\chi (\mathbf{k})$ at $\Lambda = \Lambda_\mathrm{c}$ is taken as the ordering wavevector. Plotting $\chi ( \mathbf{k} )$ at the critical cutoff over an extended region in the reciprocal space, one typically observes a rather sharp peak at $\mathbf{k}_*$, see Fig.~\ref{figure:correlatemain}l for instance. Different ordering wavevectors correspond to distinct magnetic orders and thus serve as primary labels of the latter.

However, in the parameter space of interest, we find only $\mathbf{k}=\mathbf{0}$ orders, which signify ferromagnetic correlations, when $J_x$ or $J_z$ is sufficiently negative. A further distinction is made by examining whether the $xx$ correlation at the $\Gamma$ point dominates over $zz$, or vice versa. Along the line $J_x = J_z$, we have $\chi^{xx} (\mathbf{k}) = \chi^{zz} (\mathbf{k})$ for all $\mathbf{k}$, see Fig.~\ref{figure:evolve}b for instance. Once we deviate from this line, say for $\lvert J_x \rvert < \lvert J_z \rvert$, $\chi^{zz} (\mathbf{k} = \mathbf{0})$ quickly becomes much larger than $\chi^{xx} (\mathbf{k} = \mathbf{0})$ around $\Lambda=\Lambda_\mathrm{c}$, see Fig.~\ref{figure:evolve}c for instance. In this case, the spins on the four vertices of a tetrahedron are either all aligned or all antialigned to their respective local $z$ axes, which is known as the all-in-all-out (AIAO) order. This state is labeled as Z-AIAO following the convention in Refs.~\cite{PhysRevResearch.2.023253,PhysRevLett.129.097202}, to distinguish it from the X-AIAO state where $\chi^{xx} (\mathbf{k} = \mathbf{0})$ is dominant. The Z-AIAO and X-AIAO magnetic orders are naturally separated by the $J_x = J_z$ line, see Fig.~\ref{figure:phase}.

It is much more challenging to distinguish quantum spin liquids, as they preserve the full symmetry of the Hamiltonian. While PFFRG is able to identify spin liquid ground states, it does not by itself offer a classification of different spin liquids. Nevertheless, we can try to analyze the static susceptibilities calculated from PFFRG and see if there exists any qualitative difference among them. We also make use of insights developed in other theoretical studies, which identify the $0$-flux and $\pi$-flux quantum spin ices (QSIs) as the two major spin liquid candidates in the parameter region of interest \cite{PhysRevResearch.2.023253,PhysRevB.102.104408,2301.05240}. In the perturbative regime where $J_y \gg \lvert J_x \rvert, \lvert J_z \rvert$, it has been well established \cite{PhysRevB.69.064404} that the $0$-flux ($\pi$-flux) QSI are stabilized for $(J_x+J_z) < 0$ [$(J_x+J_z) > 0$].

Hence, we plot and examine the diagonal components of $\chi^{\mu \nu} (\mathbf{k})$ in the $[hhl]$ plane in reciprocal space. Momenta are measured in the reciprocal lattice units; e.g., $(h,h,l)=(1,1,3)$ means $\mathbf{k} = 2 \pi ( \hat{\mathbf{x}} + \hat{\mathbf{y}} + 3 \hat{\mathbf{z}} ) / a$. For every parameter that stabilizes a quantum spin liquid ground state, $\chi^{yy} (\mathbf{k})$ shows bowtie-like motifs signifying spin ice correlations, with pinch point singularities at (0,0,2) and (1,1,1) that are broadened, see Figs.~\ref{figure:correlatemain}b and \ref{figure:correlatemain}f. Meanwhile, $\chi^{xx} (\mathbf{k})$ and $\chi^{zz} (\mathbf{k})$ in the spin liquid regime largely fall into two categories: both of them either show (i) diffuse peaks at the $\Gamma$ point, see Figs.~\ref{figure:correlatemain}a and \ref{figure:correlatemain}c, or (ii) the bowtie patterns similar to those in $\chi^{yy} (\mathbf{k})$ but without obvious pinch points, see Figs.~\ref{figure:correlatemain}e and \ref{figure:correlatemain}g. We associate (i) and (ii) with the $0$-flux and $\pi$-flux QSIs, respectively \cite{PhysRevLett.129.097202,2301.05240}.

These two QSIs are separated approximately by the line $J_x + J_z = 0$ (see Fig.~\ref{figure:phase}), which is consistent with perturbation theory. While there is a rather thin slice of the $0$-flux QSI in the phase diagram, the $\pi$-flux QSI occupies a much larger area, likely owing to stronger frustration from positive transverse couplings. Within the $\pi$-flux QSI, the bowtie motifs in the $xx$ correlation are sharper (more diffuse) for larger (smaller) $J_x$, and the same is true for the $zz$ correlation and $J_z$, when the color scale is chosen to range from zero to the maximum intensity. For example, one can notice the difference between Figs.~\ref{figure:correlatemain}g ($J_z=0.1$) and \ref{figure:spinflipmore}d ($J_z=0.7$).

We remark that there exists a small parameter region in which $\chi^{xx} (\mathbf{k})$ displays a maximum at the $\Gamma$ point while $\chi^{zz} (\mathbf{k})$ displays the bowtie patterns, or vice versa, see Figs.~\ref{figure:correlateorder}a, \ref{figure:correlateorder}c, \ref{figure:correlateliquid}e, and \ref{figure:correlateliquid}g in Appendix \ref{appendix:boundary}. As this only appears near the phase boundary between the $\pi$-flux QSI and the $0$-flux QSI or one of the magnetic orders, we interpret it as a tendency of the $\pi$-flux QSI to develop ferromagnetic correlations in proximity to a phase transition, and refrain from attributing it to a new kind of quantum spin liquid. Finally, we caution that it is possible for distinct quantum spin liquids to exhibit highly similar two-spin correlation functions that would not be resolved by PFFRG.

Throughout this paper, intensity plots in the $[hhl]$ plane are calculated at the smallest cutoff $\Lambda = 0.01$ for quantum spin liquids, and at the critical cutoffs $\Lambda_\mathrm{c}$ for magnetically ordered states.

\subsection{\label{section:neutron}Neutron Scattering Cross Sections}

To compute the energy-integrated neutron scattering cross section for comparisons with experiments, one should, strictly speaking, use the equal-time spin structure factor, which is formally given by the integral of the magnetic susceptibility \eqref{magneticsusceptibility} over the Mastubara frequency,
\begin{equation} \label{susceptibilityintegral}
\mathcal{S}_{ij}^\mathrm{\mu \nu} \equiv \langle S_i^\mu (0) S_j^\nu (0) \rangle = \frac{1}{2 \pi} \int \mathrm{d} \omega \, \chi_{ij}^{\mu \nu} (i \omega) .
\end{equation}
However, this additional frequency integration, which is performed over a discrete mesh, leads to further numerical errors \cite{2307.10359}. Therefore, the existing PFFRG literature mostly focuses on analyzing the static susceptibility instead of the equal-time spin correlator. We will resort to the static susceptibility \footnote{We have attempted the integration \eqref{susceptibilityintegral} with the logarithmic frequency mesh and the trapezoidal scheme mentioned in Sec.~\ref{section:numeric}, but found negative intensities in the resulting equal-time spin structure factor, which are unphysical and likely reflecting the presence of large numerical errors. A more accurate integration scheme is left as a possible future improvement.} and back our approximation with calculations using gauge mean field theory.

At very low temperatures, the momentum-resolved static susceptibility \eqref{fouriertransform} is related to the dynamical spin structure factor $\mathcal{S}^{\mu \nu} (\mathbf{k},\omega)$ via the Kramers-Kronig relation and the fluctuation-dissipation theorem \cite{colemanbook,s41467-020-15594-1,2307.10359},
\begin{equation} \label{kramerskronig}
\chi^{\mu \nu} (\mathbf{k}, i \omega=0) \propto \int \mathrm{d} \omega' \, \frac{\mathcal{S}^{\mu \nu} (\mathbf{k},\omega')}{\omega'} .
\end{equation}
On the other hand, integrating the dynamical spin structure factor over energy, one obtains the momentum-resolved equal-time spin structure factor,
\begin{equation} \label{equaltime}
\int \mathrm{d} \omega \, \mathcal{S}^{\mu \nu} (\mathbf{k},\omega) = \mathcal{S}^{\mu \nu} (\mathbf{k}) \equiv \frac{1}{N} \sum_{ij} e^{i \mathbf{k} \cdot (\mathbf{r}_i - \mathbf{r}_j)} \mathcal{S}^{\mu \nu}_{ij} .
\end{equation}
Note that $\chi^{\mu \nu} (\mathbf{k})$ and $\mathcal{S}^{\mu \nu} (\mathbf{k})$ defined above measure the two-spin correlations in the local coordinates.

To make connections with experiments, one should consider the (total) neutron scattering structure factor \begin{equation} \label{totalneutron}
S_\mathrm{TOT} (\mathbf{k}) = \frac{1}{N} \sum_{ij} \bigg[ \hat{\mathbf{z}}_i \cdot \hat{\mathbf{z}}_j - \frac{(\hat{\mathbf{z}}_i \cdot \mathbf{k}) (\hat{\mathbf{z}}_j \cdot \mathbf{k})}{\lvert \mathbf{k} \rvert^2} \bigg] e^{i \mathbf{k} \cdot (\mathbf{r}_i - \mathbf{r}_j)} \mathcal{S}^{zz}_{ij},
\end{equation}
where $\hat{\mathbf{z}}_i$ is the unit vector along the local $z$ axis at site $i$, assuming that external magnetic fields and neutrons only couple to the local $z$ components of the pseudospins. In polarized neutron scattering \eqref{totalneutron} is further decomposed into spin-flip (SF) and non-spin-flip (NSF) channels \cite{PhysRevB.86.075154,PhysRevX.12.021015,PhysRevLett.129.097202},
\begin{subequations}
\begin{align}
S_\mathrm{SF} (\mathbf{k}) &= \frac{1}{N} \sum_{ij} [ \hat{\mathbf{v}} (\mathbf{k}) \cdot \hat{\mathbf{z}}_i ] [ \hat{\mathbf{v}} (\mathbf{k}) \cdot \hat{\mathbf{z}}_j ] e^{i \mathbf{k} \cdot (\mathbf{r}_i - \mathbf{r}_j)} \mathcal{S}_{ij}^{zz} , \label{spinflip} \\
S_\mathrm{NSF} (\mathbf{k}) &= \frac{1}{N} \sum_{ij} (\hat{\mathbf{u}} \cdot \hat{\mathbf{z}}_i) (\hat{\mathbf{u}} \cdot \hat{\mathbf{z}}_j) e^{i \mathbf{k} \cdot (\mathbf{r}_i - \mathbf{r}_j)} \mathcal{S}_{ij}^{zz} , \label{nonspinflip}
\end{align}
\end{subequations}
where $\hat{\mathbf{u}}$ is the unit vector along the direction of the neutron polarization, which is perpendicular to the scattering plane, and $\hat{\mathbf{v}} (\mathbf{k}) = (\hat{\mathbf{u}} \times \mathbf{k}) / \lvert \hat{\mathbf{u}} \times \mathbf{k} \rvert$.

Due to the numerical uncertainties in calculating $\mathcal{S}_{ij}^{zz}$ as mentioned in the beginning of this subsection, we replace it by $\chi_{ij}^{zz}$ in \eqref{totalneutron}, \eqref{spinflip}, and \eqref{nonspinflip}, and denote the resulting total, spin-flip, and non-spin-flip neutron scattering structure factors by $\chi_\mathrm{TOT}(\mathbf{k})$, $\chi_\mathrm{SF} (\mathbf{k})$, and $\chi_\mathrm{NSF} (\mathbf{k})$, respectively. We then compare our results to the experimental data reported in Ref.~\cite{PhysRevX.12.021015}. While the static susceptibility is not quite the equal-time spin correlator, it is a reasonable estimate of the latter if the spectral weight of $\mathcal{S}(\mathbf{k}, \omega)$ is concentrated at low energies, or $\mathcal{S}(\mathbf{k}, \omega)$ is nonzero only within a relatively narrow range of finite energies, by \eqref{kramerskronig} and \eqref{equaltime}. We further support the replacement by a direct comparison between the static susceptibility and the equal-time spin structure factor calculated from the gauge mean field theory in Sec.~\ref{section:gmft}.

\begin{figure*}
\includegraphics[scale=0.23]{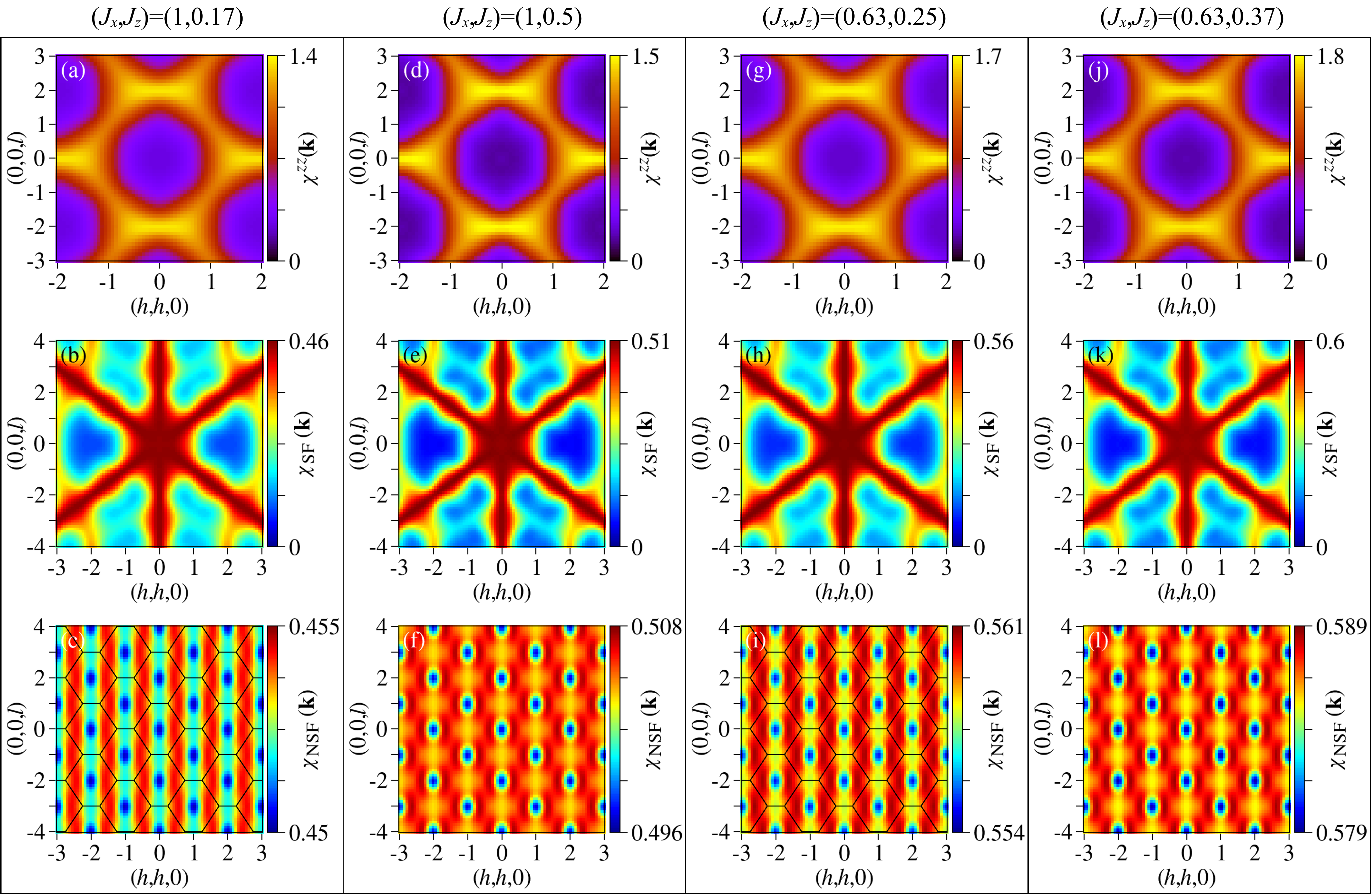}
\caption{\label{figure:spinflip}The $zz$ component of the static susceptibility, and the spin-flip and non-spin-flip neutron scattering structure factors calculated from it, in the $[hhl]$ plane, at $(J_x, J_z)$ equal to (a-c) $(1, 0.17)$, (d-f) $(1,0.5)$, (g-i) $(0.63,0.25)$, and (j-l) $(0.63,0.37)$. All these parameters stabilize the $\pi$-flux QSI. Black lines in (c) and (i) indicate the Brillouin zone boundaries of the face-centered cubic lattice.}
\end{figure*}

We first examine the two proposed parametrizations for Ce$_2$Zr$_2$O$_7$ in the existing literature. Ref.~\cite{PhysRevX.12.021015} finds $(J_x, J_y, J_z) = (0.063, 0.064, 0.011)$ meV up to a permutation of $J_x$ and $J_y$, which is supported by a subsequent work \cite{PhysRevB.108.054438}. On the other hand, Ref.~\cite{s41535-022-00458-2} finds a number of parametrizations clustered in the region $(J_x,J_y,J_z) = (0.05 \pm 0.02, 0.08 \pm 0.01, 0.02 \pm 0.01)$ meV. Scaling $J_y$ to 1, we get $(J_x,J_z) \approx (1, 0.17)$ and $(0.63,0.25)$, both of which lie within the $\pi$-flux QSI phase, see Fig.~\ref{figure:phase}. The calculated spin-flip and non-spin-flip neutron scatterings are shown in Figs.~\ref{figure:spinflip}b, \ref{figure:spinflip}c, \ref{figure:spinflip}h, and \ref{figure:spinflip}i. We find that the intensity variation of $\chi_\mathrm{NSF} (\mathbf{k})$ is generally confined to a narrow range, so the profile of $\chi_\mathrm{SF} (\mathbf{k})$ highly resembles that of $\chi_\mathrm{TOT} (\mathbf{k})$. The latter is thus not shown separately.

The most eminent feature in $\chi_\mathrm{SF} (\mathbf{k})$ is the rod-like distribution of high intensities \cite{PhysRevB.99.121102}, which is seen in the polarized neutron scattering experiment \cite{PhysRevX.12.021015} as well as reproduced in a number of theoretical calculations \cite{PhysRevX.12.021015,s41535-022-00458-2,PhysRevLett.129.097202,2301.05240}. The rods also show apparent narrowing and widening in certain sections while remaining connected, see Figs.~\ref{figure:spinflip}b and \ref{figure:spinflip}h. In particular, the vertical rod is narrowed at $(0,0,2)$ and widened at $(0,0,3)$, resembling a ``neck'' and a ``head''. The experimental data in Ref.~\cite{PhysRevX.12.021015} also features a ``neck'' at $(0,0,2)$, albeit much narrower like a pinch point. Moreover, the narrowing and widening of the rods in other directions are less severe, which is consistent with the experiment.

We now turn to $\chi_\mathrm{NSF} (\mathbf{k})$. The intersection between the Brillouin zone (BZ) boundaries of the fcc lattice and the $[hhl]$ plane is a network of edge-sharing hexagons, which looks like a honeycomb lattice compressed along one of the bond directions. We observe that the minima of $\chi_\mathrm{NSF} (\mathbf{k})$ are located at the centers of these hexagons, see Figs.~\ref{figure:spinflip}c and \ref{figure:spinflip}i, which is consistent with the experiment. However, the maxima do not take place exactly along the BZ boundaries as seen in the experiment, but rather form a stripe-like pattern.

We find that increasing $J_z$ leads to a starker contrast between the narrowed and widened sections of the rods in $\chi_\mathrm{SF} (\mathbf{k})$, as well as a more diffuse intensity background for the minima in $\chi_\mathrm{NSF} (\mathbf{k})$, see Figs.~\ref{figure:spinflip}e, \ref{figure:spinflip}f, \ref{figure:spinflip}k, and \ref{figure:spinflip}l for calculations at the parameters $(J_x,J_z) = (1,0.5)$ and $(0.63,0.37)$. These changes coincide with the sharpening of the bowtie motifs in $\chi^{zz} (\mathbf{k})$: tighter ``knot'' of the bowtie is accompanied by narrower ``neck'' of the rods, c.f.~Figs.~\ref{figure:spinflip}a and \ref{figure:spinflip}d. While introducing further nearest neighbor interactions may substantially improve the agreement between theory and experiment, as demonstrated by Ref.~\cite{s41535-022-00458-2}, we have shown that the nearest neighbor XYZ model is able to qualitatively capture the main features of the polarized neutron scattering cross-sections reported in Ref.~\cite{PhysRevX.12.021015}.

\begin{figure*}
\includegraphics[scale=0.23]{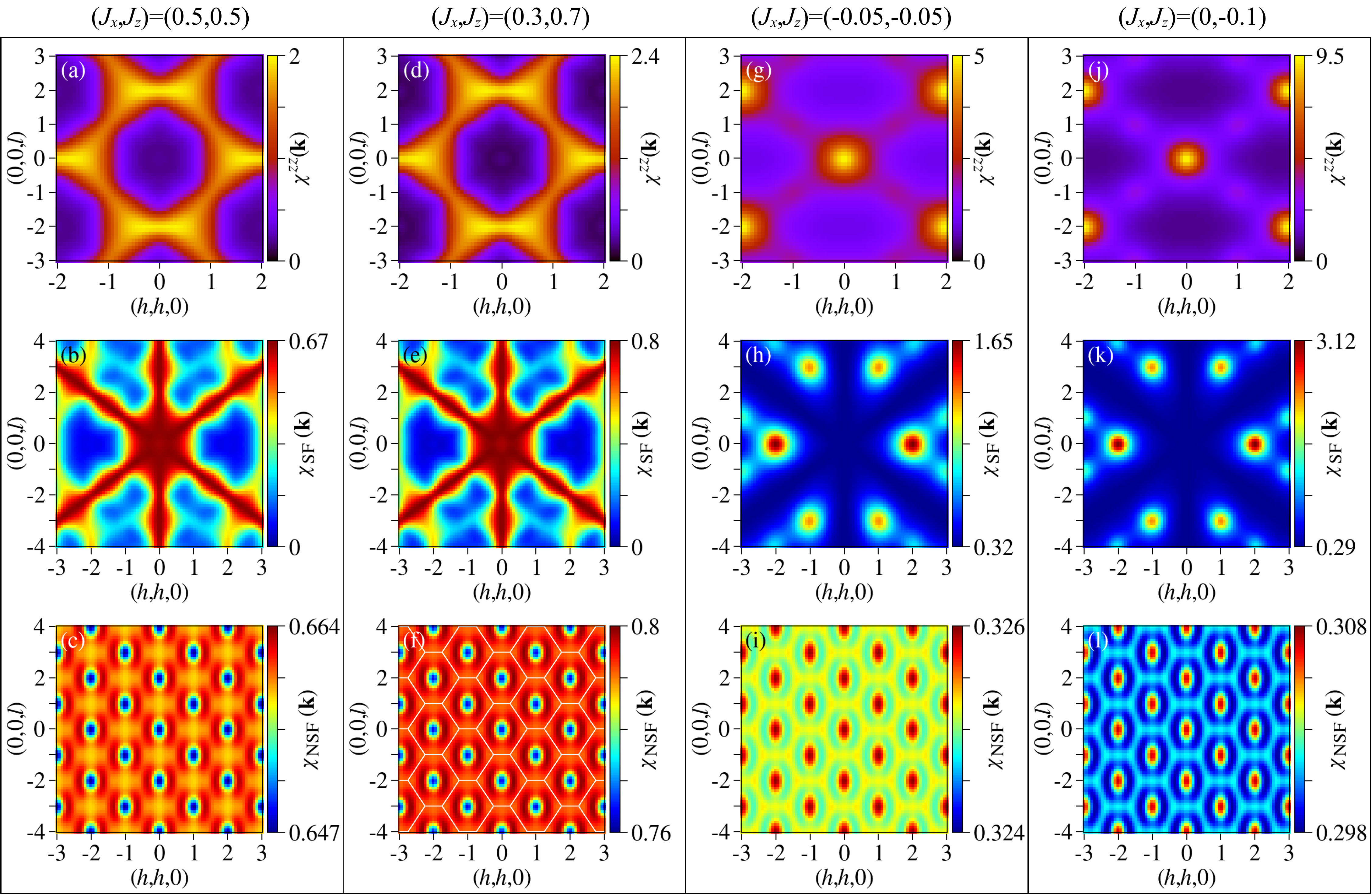}
\caption{\label{figure:spinflipmore}The $zz$ component of the static susceptibility, and the spin-flip and non-spin-flip neutron scattering structure factors calculated from it, in the $[hhl]$ plane, at $(J_x, J_z)$ equal to (a-c) $(0.5,0.5)$, (d-f) $(0.3,0.7)$, (g-i) $(-0.05,-0.05)$, and (j-l) $(0,-0.1)$. The first (last) two of these parameters stabilize the $\pi$-flux ($0$-flux) QSI. White lines in (f) indicate the Brillouin zone boundaries of the face-centered cubic lattice.}
\end{figure*}

Additional calculations of the spin-flip and non-spin-flip neutron scatterings at other parameters reveal that the $\pi$-flux QSI with $J_z \gtrsim 0.3$ generically displays both (i) high intensity rods with the head-and-neck features in the spin-flip channel, and (ii) well defined minima at the BZ centers in the non-spin-flip channel, see Figs.~\ref{figure:spinflipmore}b, \ref{figure:spinflipmore}c, \ref{figure:spinflipmore}e, and \ref{figure:spinflipmore}f for instance. For smaller or negative values of $J_z$, we can still observe the rod motifs in the spin-flip channel, but the intensity profile of the non-spin-flip channel becomes more stripe-like, which looks like Fig.~\ref{figure:spinflip}c.

Finally, we show plots of $\chi_\mathrm{SF} (\mathbf{k})$ and $\chi_\mathrm{NSF} (\mathbf{k})$ calculated at two choices of parameters that stabilize the $0$-flux QSI, see Figs.~\ref{figure:spinflipmore}h, \ref{figure:spinflipmore}i, \ref{figure:spinflipmore}k, and \ref{figure:spinflipmore}l. Roughly speaking, the intensity distribution of the $0$-flux QSI is opposite to that of the $\pi$-flux QSI: (i) the rods now carry low instead of high intensities in $\chi_\mathrm{SF} (\mathbf{k})$, while (ii) the BZ centers are now maxima instead of minima in $\chi_\mathrm{NSF} (\mathbf{k})$. These distinctions, which are also observed in two other theoretical studies \cite{PhysRevLett.129.097202,2301.05240}, can serve as criteria in differentiating the $0$-flux and $\pi$-flux QSIs in future neutron scattering experiments.

\subsection{\label{section:gmft}Gauge Mean Field Theory}

When computing the neutron scattering cross sections in Sec.~\ref{section:neutron}, we have replaced the equal-time spin correlator with the static susceptibility, as PFFRG is unable to calculate the former accurately. Here, we support the validity of this replacement by calculating and comparing the static susceptibility and the equal-time spin structure factor using gauge mean field theory (GMFT) \cite{PhysRevB.107.064404,2301.05240}.

\begin{figure*}
\includegraphics[scale=0.23]{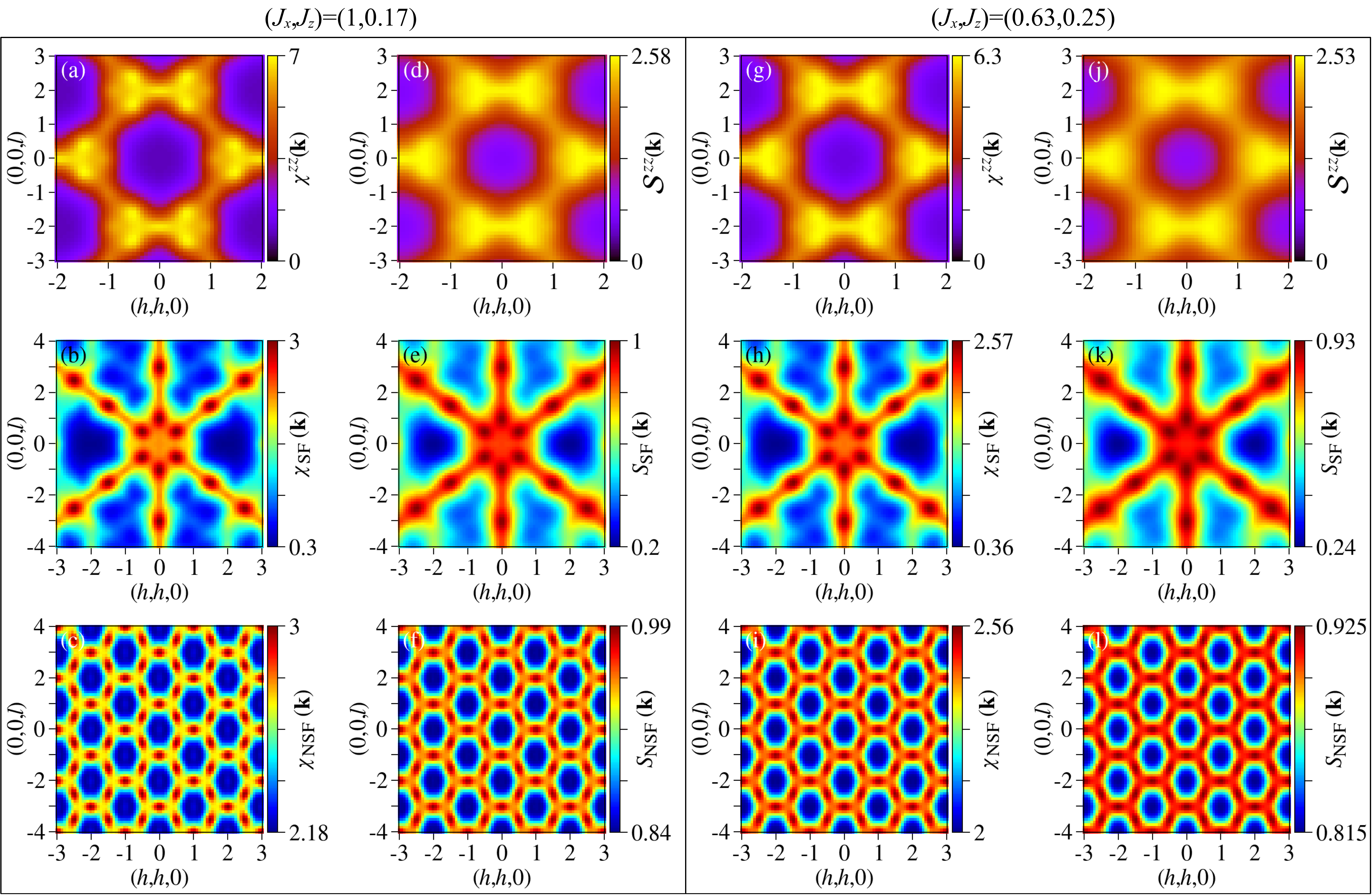}
\caption{\label{figure:gmft}The $zz$ component of the static susceptibility, and the spin-flip and non-spin-flip neutron scattering structure factors resulting from it, at $(J_x, J_z)$ equal to (a-c) $(1,0.17)$ and (g-i) $(0.63,0.25)$, calculated using gauge mean field theory. The $zz$ component of the equal-time spin structure factor, and the spin-flip and non-spin-flip neutron scattering structure factors resulting from it, at $(J_x, J_z)$ equal to (d-f) $(1,0.17)$ and (j-l) $(0.63,0.25)$, calculated using gauge mean field theory. Both parameters stabilize the $\pi$-flux QSI.}
\end{figure*}

Figs.~\ref{figure:gmft}a and \ref{figure:gmft}d show the $zz$ components of the static susceptibility and the equal-time spin structure factor at $(J_x,J_z)=(1,0.17)$. One can see that their intensity distributions are highly similar, though the profile of $\mathcal{S}^{zz} (\mathbf{k})$ appears more diffuse than $\chi^{zz} (\mathbf{k})$. Figs.~\ref{figure:gmft}b and \ref{figure:gmft}c further show the spin-flip and non-spin-flip neutron scatterings calculated from the static susceptibility, as we did with PFFRG in Sec.~\ref{section:neutron}. One can again note their similarities to Figs.~\ref{figure:gmft}e and \ref{figure:gmft}f, which are calculated from the equal-time spin structure factor directly, i.e., according to \eqref{spinflip} and \eqref{nonspinflip}. The same statements can be made in regards to $\chi^{zz} (\mathbf{k})$ and $\mathcal{S}^{zz} (\mathbf{k})$ at $(J_x,J_z)=(0.63,0.25)$, as well as the corresponding $\chi_\mathrm{(N)SF} (\mathbf{k})$ and $S_\mathrm{(N)SF} (\mathbf{k})$, when one compares Figs.~\ref{figure:gmft}g-\ref{figure:gmft}i with Figs.~\ref{figure:gmft}j-\ref{figure:gmft}l. We show additional plots of $\chi^{zz} (\mathbf{k})$ and $\mathcal{S}^{zz} (\mathbf{k})$ calculated at other parameters in Appendix \ref{appendix:gmft}, see Figs.~\ref{figure:gmftmore}a-\ref{figure:gmftmore}h. The resemblance between the static susceptibility and the equal-time spin structure factor means that the former is a good approximation of the latter.

We can further compare the GMFT results in this subsection with the PFFRG results in the previous subsection. Both show the bowtie patterns in $\chi^{zz} (\mathbf{k})$, the rod-like distributions of high intensities with the head-and-neck features in $\chi_\mathrm{SF} (\mathbf{k})$, and the minima at the BZ centers in $\chi_\mathrm{NSF} (\mathbf{k})$. In addition, GMFT reveals (i) sharp point-like maxima along the high-intensity rods in the spin-flip channel, so that an apparent dent of intensity is seen in the vicinity of $\mathbf{k}=\mathbf{0}$, as well as (ii) maxima along the BZ boundaries in the non-spin-flip channel. The agreement between the GMFT results and the experimental data in Ref.~\cite{PhysRevX.12.021015} is excellent.

\section{\label{section:discussion}Discussion}

In summary, we have employed the pseudofermion functional renormalization group (PFFRG) to study the nearest neighbor XYZ model on the pyrochlore lattice, in the parameter regime relevant to the quantum spin liquid candidate Ce$_2$Zr$_2$O$_7$. PFFRG analyses of pyrochlore magnets in the existing literature have mostly focused on Heisenberg models \cite{PhysRevMaterials.1.071201,PhysRevX.9.011005,PhysRevB.105.054426,SciPostPhys.12.5.156,PhysRevResearch.4.023185}, which are isotropic in spin space. Applications of PFFRG to anisotropic spin models on the pyrochlore lattice only appeared quite recently; our work contributes to this effort in addition to two others that investigate the Heisenberg-Dzyaloshinskii-Moriya model \cite{PhysRevB.107.214414} and the non-Kramers pyrochlore model \cite{2310.16682}.

We present a phase diagram that contains two quantum spin ices (QSIs) and two magnetically ordered states, and the phase boundaries largely agree with existing theoretical works \cite{PhysRevResearch.2.023253,PhysRevB.102.104408,2301.05240}. Approximating the equal-time spin structure factor by the static susceptibility, we compute the spin-flip and non-spin-flip channels of the neutron scattering cross-sections at various parameters. We back such an approximation with calculations using gauge mean field theory (GMFT) \cite{PhysRevB.107.064404,2301.05240}. We find that the computed neutron scattering cross-sections are able to reproduce several qualitative features seen in the experimental data of Ref.~\cite{PhysRevX.12.021015} across a wide range of parameters within the $\pi$-flux QSI phase. In other words, we have demonstrated that a reasonable agreement with the neutron scattering experiment can already be obtained at the level of nearest neighbor interactions, though our results may be further refined by including second nearest neighbor interactions as proposed in Ref.~\cite{s41535-022-00458-2}. More importantly, our results support the case of a quantum spin liquid ground state in Ce$_2$Zr$_2$O$_7$, which is likely the $\pi$-flux QSI.

Apart from PFFRG, theoretical methods such as numerical linked cluster \cite{PhysRevX.12.021015}, molecular dynamics \cite{PhysRevLett.129.097202}, exact diagonalization \cite{PhysRevLett.129.097202}, and gauge mean field theory \cite{2301.05240} are able to reproduce the rod motifs seen in the spin-flip channel. However, numerical linked cluster and molecular dynamics do not capture the intensity variation in the non-spin-flip neutron channel, while exact diagonalization and gauge mean field theory do. GMFT currently yields the best agreement with the polarized neutron scattering experiment \cite{PhysRevX.12.021015} among these methods. A rather intriguing feature of the GMFT calculations is that the mean field amplitudes associated with the non-$S^y$-conserving terms converge to zero, which effectively reduces the XYZ model to an XYX model \cite{2301.05240}. Such an emergent $U(1)$ symmetry is corroborated by exact diagonalization results \cite{PhysRevLett.129.097202}, which find $\mathcal{S}^{xx} (\mathbf{k}) = \mathcal{S}^{zz} (\mathbf{k})$ even though $J_x \neq J_z$. On the other hand, molecular dynamics \cite{PhysRevLett.129.097202} and the PFFRG analysis in this work predict that $\mathcal{S}^{xx} (\mathbf{k}) \neq \mathcal{S}^{zz} (\mathbf{k})$ and $\chi^{xx} (\mathbf{k}) \neq \chi^{zz} (\mathbf{k})$ for $J_x \neq J_z$. It will be interesting to resolve this disagreement in future investigations.

We also point out several other possible directions for future studies. As mentioned in the main text, the pseudofermion representation of spins in PFFRG introduces unphysical states with zero or double occupancies, which are argued to be thermally suppressed in the zero temperature limit. However, they may become important at finite temperatures, which renders the PFFRG application inaccurate. To overcome this problem, the pseudo-Majorana functional renormalization group (PMFRG) \cite{PhysRevB.103.104431,SciPostPhys.12.5.156,PhysRevLett.130.196601} has recently been developed, where the spin operator is represented in terms of SO(3) Majorana fermions. This construction generates no unphysical states, but rather redundant physical states due to a $\mathbb{Z}_2$ gauge freedom. PMFRG allows the calculations of thermodynamic quantities such as free energy and heat capacity as a function of temperature. It will be interesting to apply PMFRG to the pyrochlore XYZ model. For the truncation of the infinite hierarchy of FRG integro-differential equations, one may also apply the more elaborate multiloop scheme \cite{PhysRevResearch.4.023185,2011.01268} rather than the Katanin scheme.

From a broader perspective, it will be desirable to obtain the dynamical spin structure factor directly from PFFRG. This is not possible with the current PFFRG scheme, which is formulated in imaginary time, as the analytic continuation from Matsubara to real frequencies is a challenging numerical problem \cite{reutherthesis,PhysRevE.94.023303,buessenthesis}. The ability to calculate the dynamical spin structure factor would make PFFRG a more powerful theoretical tool in the study of frustrated magnetism, given the importance of inelastic neutron scattering experiments.

\begin{acknowledgements}
For the purpose of open access, the author has applied a Creative Commons Attribution (CC BY) licence to any Author Accepted Manuscript version arising from this submission. We thank Finn Lasse Buessen and Dominik Kiese for discussions on the PFFRG method. CC and LEC were supported by Engineering and Physical Sciences Research Council grants No.~EP/T028580/1 and No.~EP/V062654/1. YBK and FD were supported by the NSERC of Canada and the Centre for Quantum Materials at the University of Toronto. YBK was further supported by the Simons Fellowship from the Simons Foundation and the Guggenheim Fellowship from the John Simon Guggenheim Memorial Foundation. FD was further supported by the Vanier Canada Graduate Scholarship.
\end{acknowledgements}

\appendix

\section{\label{appendix:pffrg}Pseudofermion Functional Renormalization Group}

We provide additional details of the PFFRG method following \eqref{barepropagator} in Sec.~\ref{section:pffrg} of the main text. For notational simplicity, we hide the imaginary unit associated with the Matsubara frequency when it appears as the argument of a function, e.g., we write the bare propagator as $G_0^{\Lambda} (\omega)$ instead of $G_0^{\Lambda} (i \omega)$, in this appendix.

Introducing the infrared cutoff $\Lambda$, the full propagator becomes
\begin{equation} \label{fullpropagator}
G^\Lambda (\omega) = \frac{ \theta (\lvert \omega \rvert - \Lambda)}{i \omega - \Sigma^\Lambda (\omega) } ,
\end{equation}
where $\Sigma (\omega)$ is the self-energy. We also define the single-scale propagator
\begin{equation} \label{singlescalepropagator}
S^\Lambda (\omega) = [ G^\Lambda (\omega) ]^2 \frac{ \mathrm{d} [ G_0^\Lambda (\omega) ]^{-1} }{\mathrm{d} \Lambda} = \frac{ \delta ( \lvert \omega \rvert - \Lambda ) }{ i \omega - \Sigma^\Lambda (\omega) } .
\end{equation}

In the Katanin truncation scheme \cite{PhysRevB.70.115109}, all vertices with $n \geq 3$ are neglected, while each single-scale propagator \eqref{singlescalepropagator} appearing in the flow of the $n=2$ vertex is replaced by
\begin{equation} \label{kataninpropagator}
S_\mathrm{kat}^\Lambda ( \omega ) = S^\Lambda ( \omega ) - [G^\Lambda (\omega)]^2 \frac{ \mathrm{d} \Sigma^\Lambda ( \omega ) }{ \mathrm{d} \Lambda }
\end{equation}
as an attempt to partially recover the contribution of the discarded $n=3$ vertex \cite{reutherthesis,buessenthesis}. It has been phenomenologically demonstrated that without the above substitution, PFFRG is unable to capture paramagnetic ground states such as quantum spin liquids \cite{PhysRevB.81.144410,2307.10359,buessenthesis}.

\begin{figure*}
\includegraphics[scale=0.48]{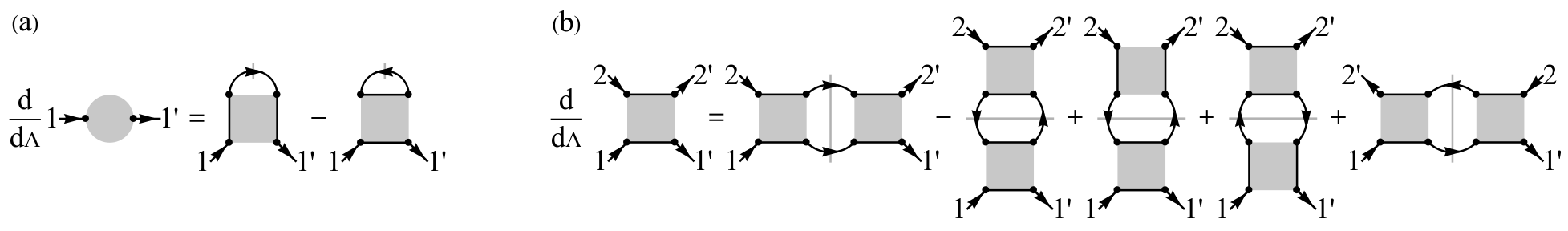}
\caption{\label{figure:flow}Diagrammatic representations of the flows of (a) self-energy and (b) two-particle vertex, see \eqref{selfenergyflow} and \eqref{vertexflow}. The slashed propagator in (a) represents the single-scale propagator $S^\Lambda (\omega)$ defined in \eqref{singlescalepropagator}, while the pair of slashed propagators in (b) represents the sum of the two cases where one of the propagators is replaced by $S^\Lambda_\mathrm{kat} (\omega)$ defined in \eqref{kataninpropagator}. Site indices are conserved along solid lines. This figure is adapted from Ref.~\cite{2307.10359}.}
\end{figure*}

To simplify the notation, we use the index $n \in \mathbb{N}$ to denote the $3$-tuple $(i_n, \alpha_n, \omega_n)$ of lattice site, spin flavor, and Matsubara frequency. An intermediate parametrization of the two-particle vertex reads
\begin{equation}
\begin{split}
\Gamma^\Lambda (1' , 2' ; 1 , 2) &= \Gamma^\Lambda_{i_1 i_2} (1', 2' ; 1 , 2) \delta_{i_{1'} i_1} \delta_{i_{2'} i_2} \\ & \quad - \Gamma^\Lambda_{i_1 i_2} (2' , 1' ; 1 , 2) \delta_{i_{1'} i_2} \delta_{i_{2'} i_1} ,
\end{split}
\end{equation}
where the primed (unprimed) indices label outgoing (incoming) fermions, and the second term differ from the first one by a crossing. The flow equations of the self-energy and the two-particle vertex then read \cite{buessenthesis}
\begin{widetext}
\begin{subequations}
\begin{align}
& \frac{\mathrm{d}}{\mathrm{d} \Lambda} \Sigma^\Lambda (\omega_1) = \frac{1}{2 \pi} \sum_{\alpha_2 \omega_2} \big[ \Gamma^\Lambda_{i_1 i_1} (2,1;1,2) - \sum_j \Gamma^\Lambda_{i_1 j} (1,2;1,2) \big] S^\Lambda (\omega_2) , \label{selfenergyflow} \\
\begin{split} \label{vertexflow}
& \frac{\mathrm{d}}{\mathrm{d} \Lambda} \Gamma^\Lambda (1',2';1,2) = \frac{1}{2 \pi} \sum_{\substack{\alpha_3 \omega_3 \\ \alpha_4 \omega_4}} \big[ \Gamma^\Lambda_{i_1 i_2} (1',2';3,4) \Gamma^\Lambda_{i_1 i_2} (3,4;1,2) - \sum_j \Gamma^\Lambda_{i_1 j} (1',4;1,3) \Gamma^\Lambda_{j i_2} (3,2';4,2) + \Gamma^\Lambda_{i_1 i_2} (1',4;1,3) \\
& \times \Gamma^\Lambda_{i_2 i_2} (2',3;4,2) + \Gamma^\Lambda_{i_1 i_1} (4,1';1,3) \Gamma^\Lambda_{i_1 i_2} (3,2';4,2) + \Gamma^\Lambda_{i_1 i_2} (2',4;1,3) \Gamma^\Lambda_{i_1 i_2} (3,1';4,2) \big] \big[ G^\Lambda (\omega_3) S^\Lambda_\mathrm{kat} (\omega_4) + (3 \longleftrightarrow 4) \big] ,
\end{split}
\end{align}
\end{subequations}
\end{widetext}
which are subjected to the initial conditions
\begin{subequations}
\begin{align}
& \Sigma (\omega) \big \vert_{\Lambda \longrightarrow \infty} = 0 , \\
& \Gamma_{i_1 i_2} (1',2';1,2) \big \vert_{\Lambda \longrightarrow \infty} = \frac{J_{i_1 i_2}^{\mu \nu}}{4} \sigma_{\alpha_{1'} \alpha_1}^\mu \sigma_{\alpha_{2'} \alpha_2}^\nu .
\end{align}
\end{subequations}
The flow equations are diagrammatically represented in Figs.~\ref{figure:flow}a and \ref{figure:flow}b. Exploiting the full symmetry of the problem, the expressions of \eqref{selfenergyflow} and \eqref{vertexflow} become more elaborate, see \cite{SciPostPhysCodeb.5} for instance.

The magnetic susceptibility \eqref{magneticsusceptibility} is calculated using the two-particle vertex and the full propagator as \cite{negeleorlandbook,2307.10359,buessenthesis}
\begin{widetext}
\begin{equation}
\begin{aligned}[b]
\chi_{ij}^{\mu \nu} (\omega) &= - \frac{1}{4 \pi} \int \mathrm{d} \omega_1 \, G^\Lambda (\omega_1) G^\Lambda (\omega_1 + \omega) \delta_{ij} \delta_{\mu \nu} \\
& \quad - \frac{1}{16 \pi^2} \int \mathrm{d} \omega_1 \int \mathrm{d} \omega_2 \, G^\Lambda (\omega_1 + \omega)  G^\Lambda (\omega_1) G^\Lambda (\omega_2) G^\Lambda (\omega_2 + \omega) \sum_{\substack{\alpha_{1'} \alpha_1 \\ \alpha_{2'} \alpha_2}} \Gamma^\Lambda (1',2';1,2) \sigma^\mu_{\alpha_1 \alpha_{1'}} \sigma^\nu_{\alpha_2 \alpha_{2'}} \\
& \includegraphics[scale=0.56]{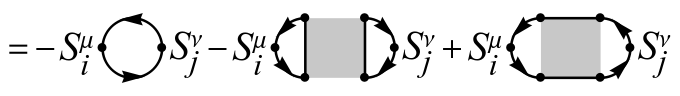}
\end{aligned}
\end{equation}
\end{widetext}
where $1'=(i,\alpha_{1'},\omega_1+\omega)$, $2'=(j,\alpha_{2'},\omega_2)$, $1=(i,\alpha_1,\omega_1)$, and $2=(j,\alpha_2,\omega_2+\omega)$.

\section{\label{appendix:boundary}Nearing Phase Boundaries}

We plot the diagonal components of the static susceptibility and their summation in the $[hhl]$ plane, at several parameters in the vicinity of the phase boundaries.

\begin{figure*}
\includegraphics[scale=0.23]{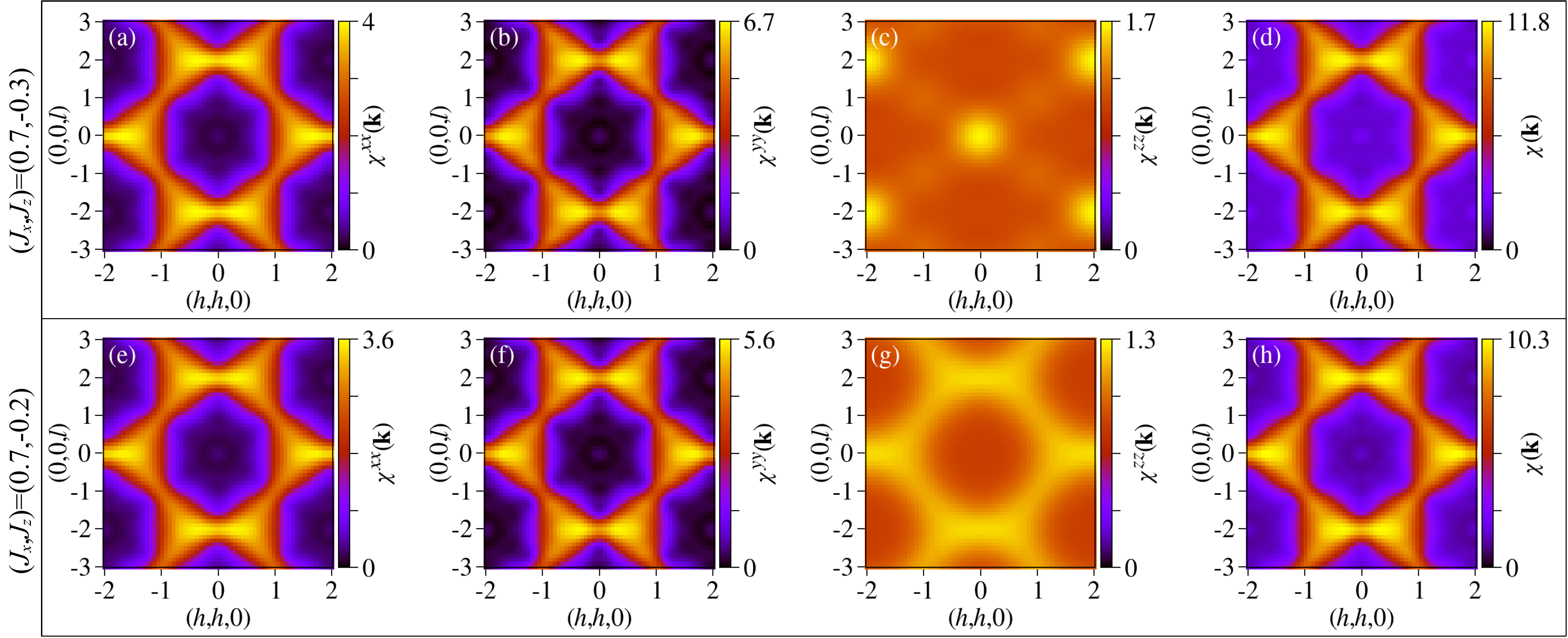}
\caption{\label{figure:correlateorder}Diagonal components of the static susceptibility, as well as their summation, in the $[hhl]$ plane, at $(J_x, J_z)$ equal to (a-d) $(0.7, -0.3)$ and (e-h) $(0.7,-0.2)$, which are located in the $\pi$-flux QSI phase.}
\end{figure*}

\begin{figure*}
\includegraphics[scale=0.23]{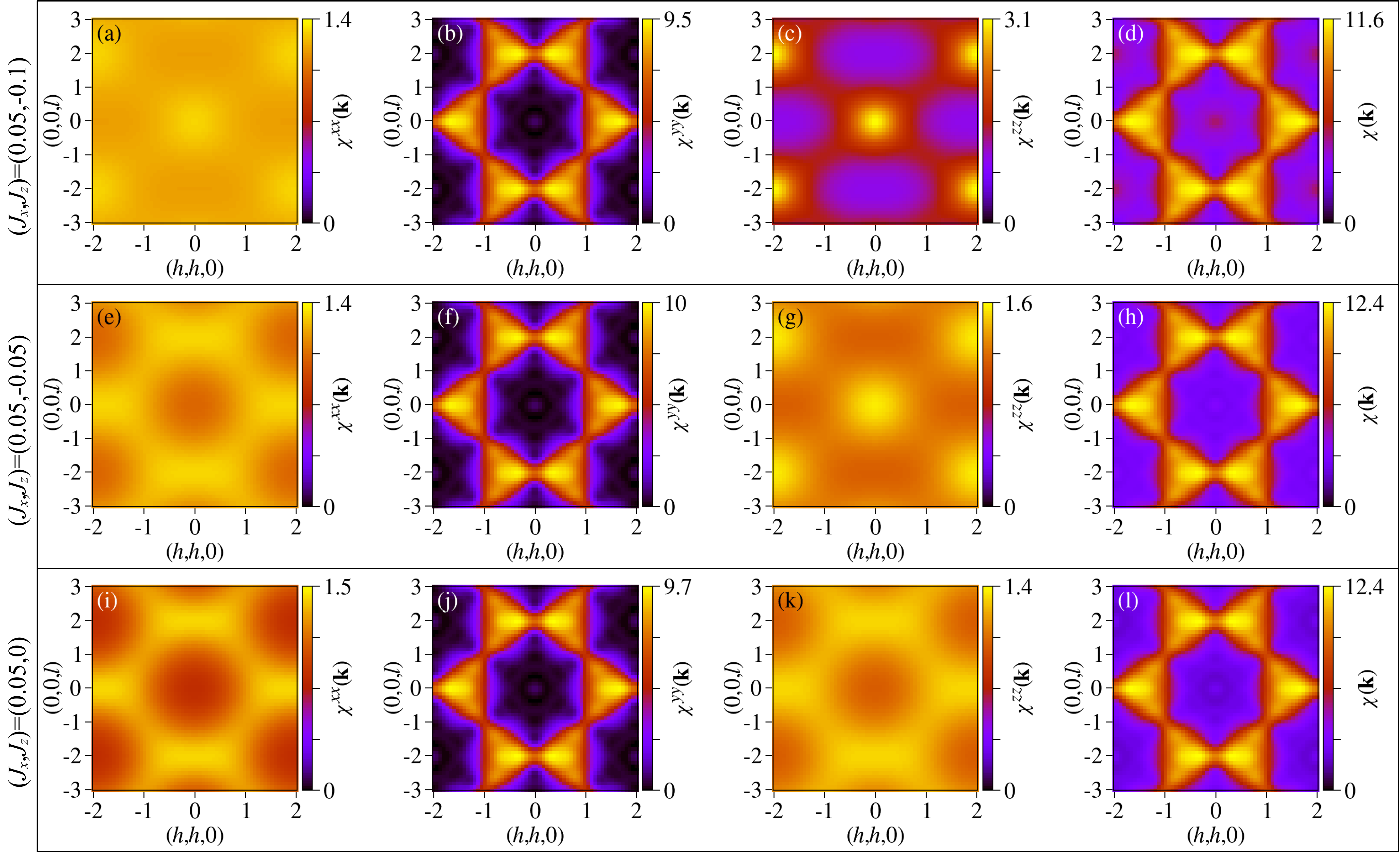}
\caption{\label{figure:correlateliquid}Diagonal components of the static susceptibility, as well as their summation, in the $[hhl]$ plane, at $(J_x, J_z)$ equal to (a-d) $(0.05, -0.1)$, (e-h) $(0.05,-0.05)$, and (i-l) $(0.05,0)$, which are located in the $0$-flux QSI phase, right at the phase boundary between the two QSIs, and in the $\pi$-flux QSI phase, respectively.}
\end{figure*}

At $(J_x, J_z) = (0.7,-0.3)$, the system is within the $\pi$-flux QSI phase but close to the phase transition into the Z-AIAO magnetic order. As a result, a small and diffuse peak appears at the $\Gamma$ point in the $zz$ correlation, signifying a buildup of weak ferromagnetic correlations, while the $xx$ correlation clearly shows the bowtie patterns, see Figs.~\ref{figure:correlateorder}a and \ref{figure:correlateorder}c. Upon increasing $J_z$ to $-0.2$, the $\Gamma$ peak is replaced by the bowtie patterns in the $zz$ correlation despite $J_z$ still being negative, though the intensity profile appears more diffuse than that of the $xx$ correlation, see Figs.~\ref{figure:correlateorder}e and \ref{figure:correlateorder}g.

At $(J_x, J_z) = (0.05,-0.05)$, the system is located right at the phase boundary between the two QSIs. The $zz$ correlation shows a small and diffuse peak at the $\Gamma$ point, while the $xx$ correlation shows diffuse bowtie patterns with a minimum at the $\Gamma$ point, see Figs.~\ref{figure:correlateliquid}e and \ref{figure:correlateliquid}g. Moving slightly away from the phase boundary towards the 0-flux QSI or the $\pi$-flux QSI, one observes that both the $xx$ and $zz$ correlations either display the $\Gamma$ peaks or the bowtie patterns, i.e.~they agree in the overall distribution of intensities, see Figs.~\ref{figure:correlateliquid}a, \ref{figure:correlateliquid}c, \ref{figure:correlateliquid}i, and \ref{figure:correlateliquid}k. It is also worth noting that the $yy$ correlation remains dominant and almost unchanged as $J_z$ is increased from $-0.1$ to $0$, since $J_y$ is much larger in magnitude than $J_x$ and $J_z$.

\section{\label{appendix:gmft}Gauge Mean Field Theory}

\begin{figure*}
\includegraphics[scale=0.23]{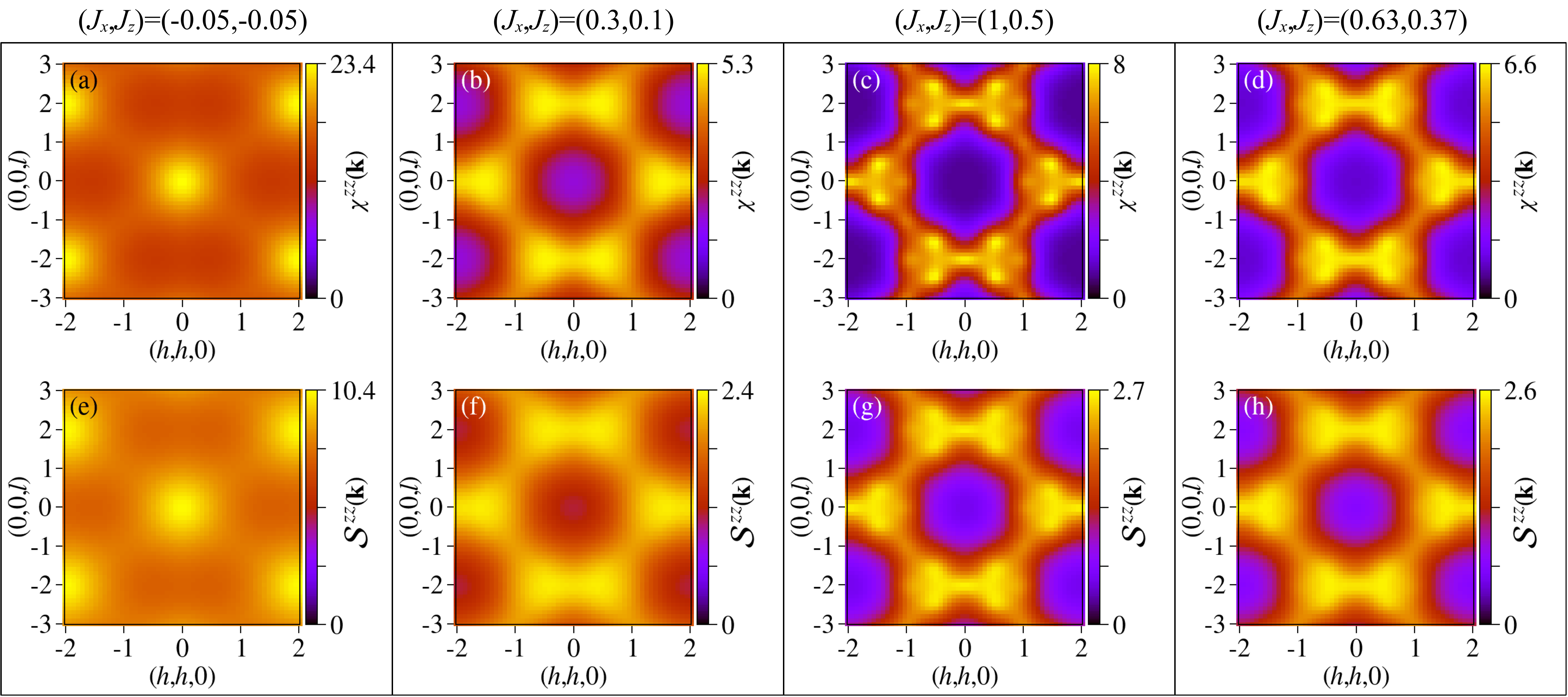}
\caption{\label{figure:gmftmore}The $zz$ component of the static susceptibility at $(J_x,J_y)$ equal to (a) $(-0.05,-0.05)$, (b) $(0.3,0.1)$, (c) $(1,0.5)$, and (d) $(0.63,0.37)$, in the $[hhl]$ plane, calculated by gauge mean field theory. The $zz$ component of the equal-time spin structure factor at $(J_x,J_y)$ equal to (e) $(-0.05,-0.05)$, (f) $(0.3,0.1)$, (g) $(1,0.5)$, and (h) $(0.63,0.37)$, in the $[hhl]$ plane, calculated by gauge mean field theory.}
\end{figure*}

In Figs.~\ref{figure:gmftmore}a-\ref{figure:gmftmore}h, we show the static susceptibilities and the equal-time spin structure factors calculated by GMFT at various parameters, for further comparisons. Note the similarities of the intensity distributions, and that the profile of $\mathcal{S}^{zz} (\mathbf{k})$ is generally more diffuse than that of $\chi^{zz} (\mathbf{k})$.

\section{\label{appendix:width}Interpolating between Ising Limits}

In this appendix, we investigate the XYZ model \eqref{xyzhamiltonian} in the parameter regime $J_x, J_y, J_z \geq 0$ without the restriction that $J_y$ is dominant. When $J_z \gg J_x,J_y$, we have a $\pi$-flux quantum spin ice where the majority of tetrahedra obeys the 2I2O ice rule along to the local $z$ axes, which we call the $z$-QSI. Increasing the strength of $J_{x,y}$ enhances quantum fluctuations that allow the creation of $z$-monopoles. The $x$-QSI and $y$-QSI together with their monopoles are similarly defined. We would like to understand how the $x$-, $y$-, and $z$-QSIs are related as we traverse the parameter space from one Ising limit to the other.

\begin{figure}
\includegraphics[scale=0.2]{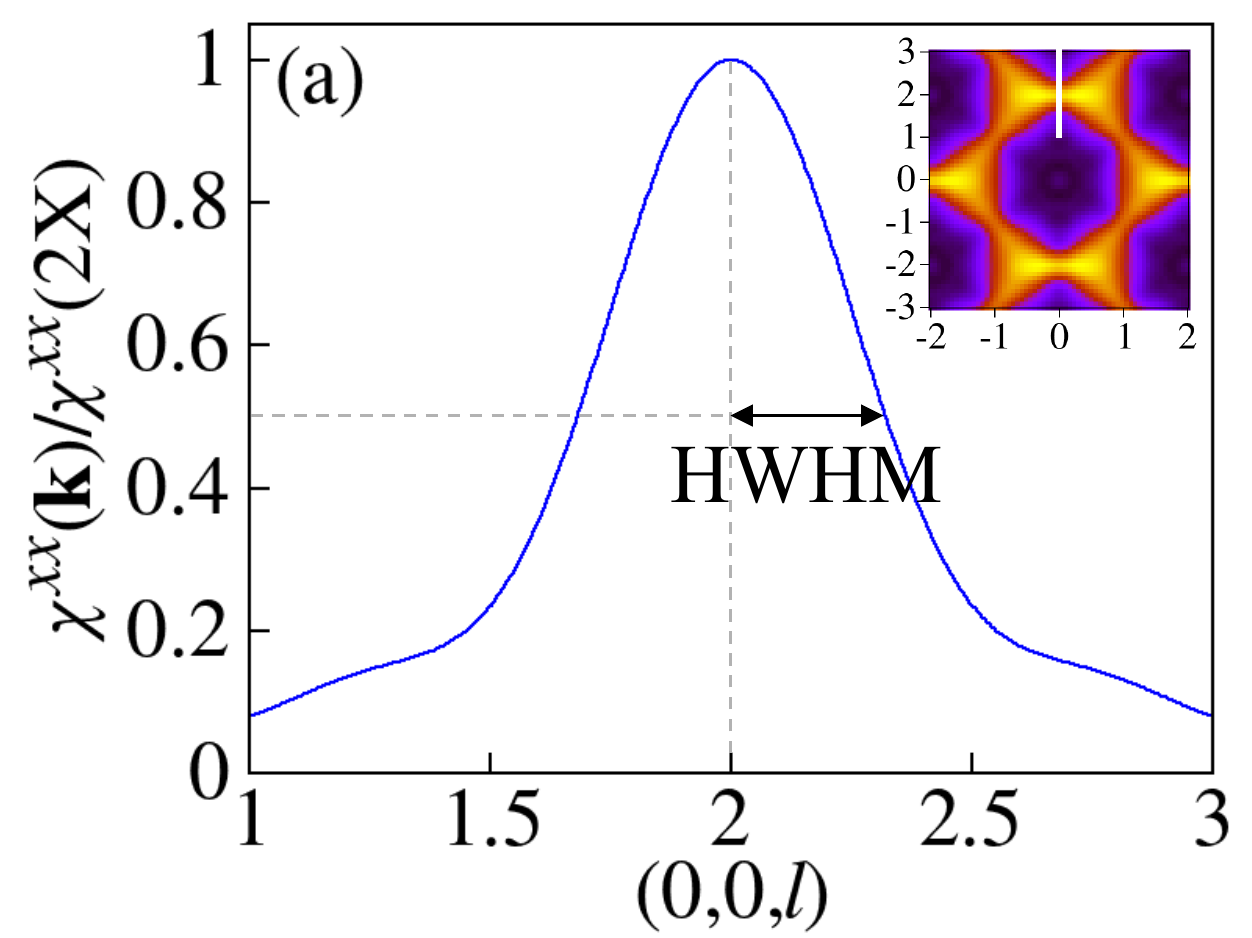}
\includegraphics[scale=0.2]{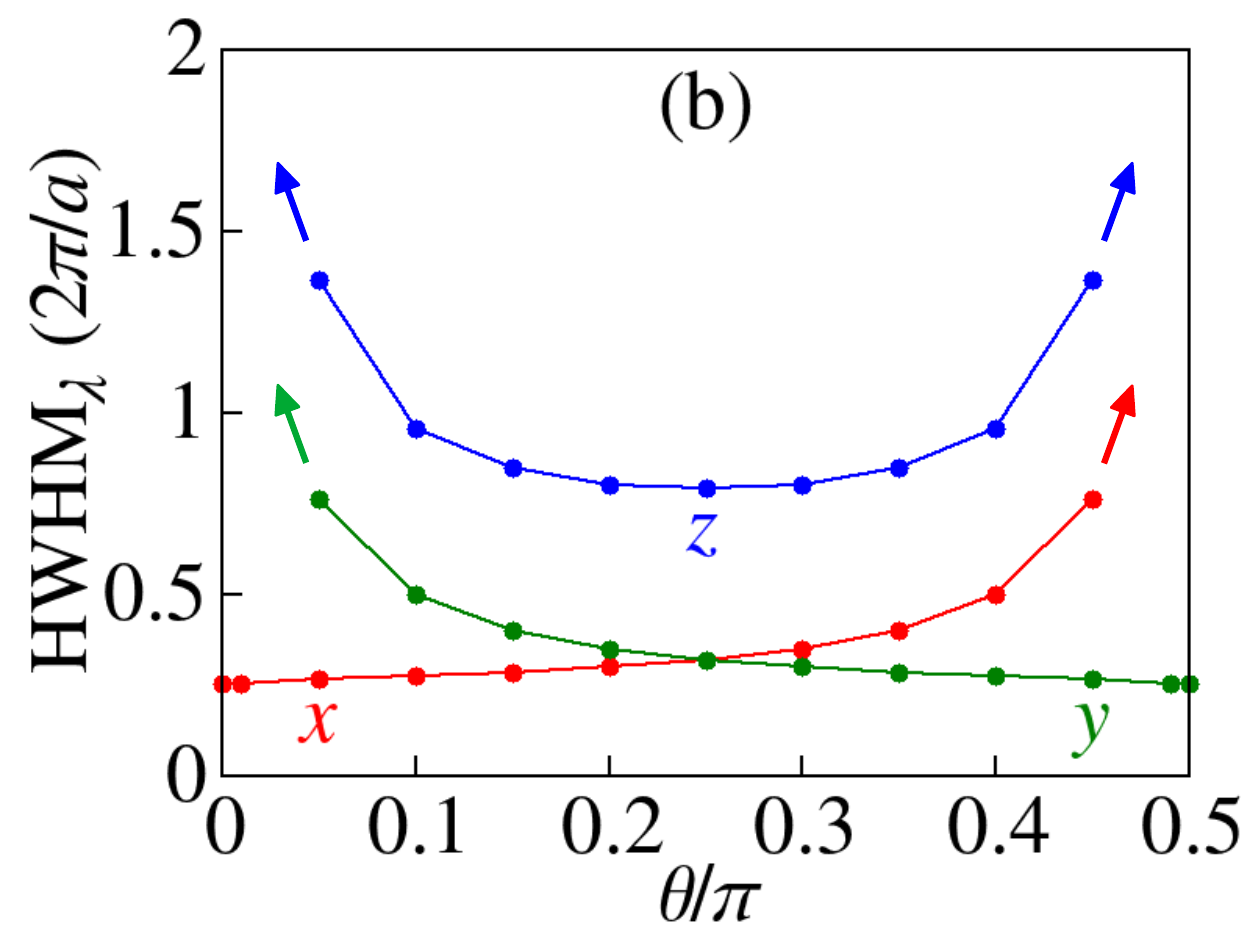} \\
\includegraphics[scale=0.2]{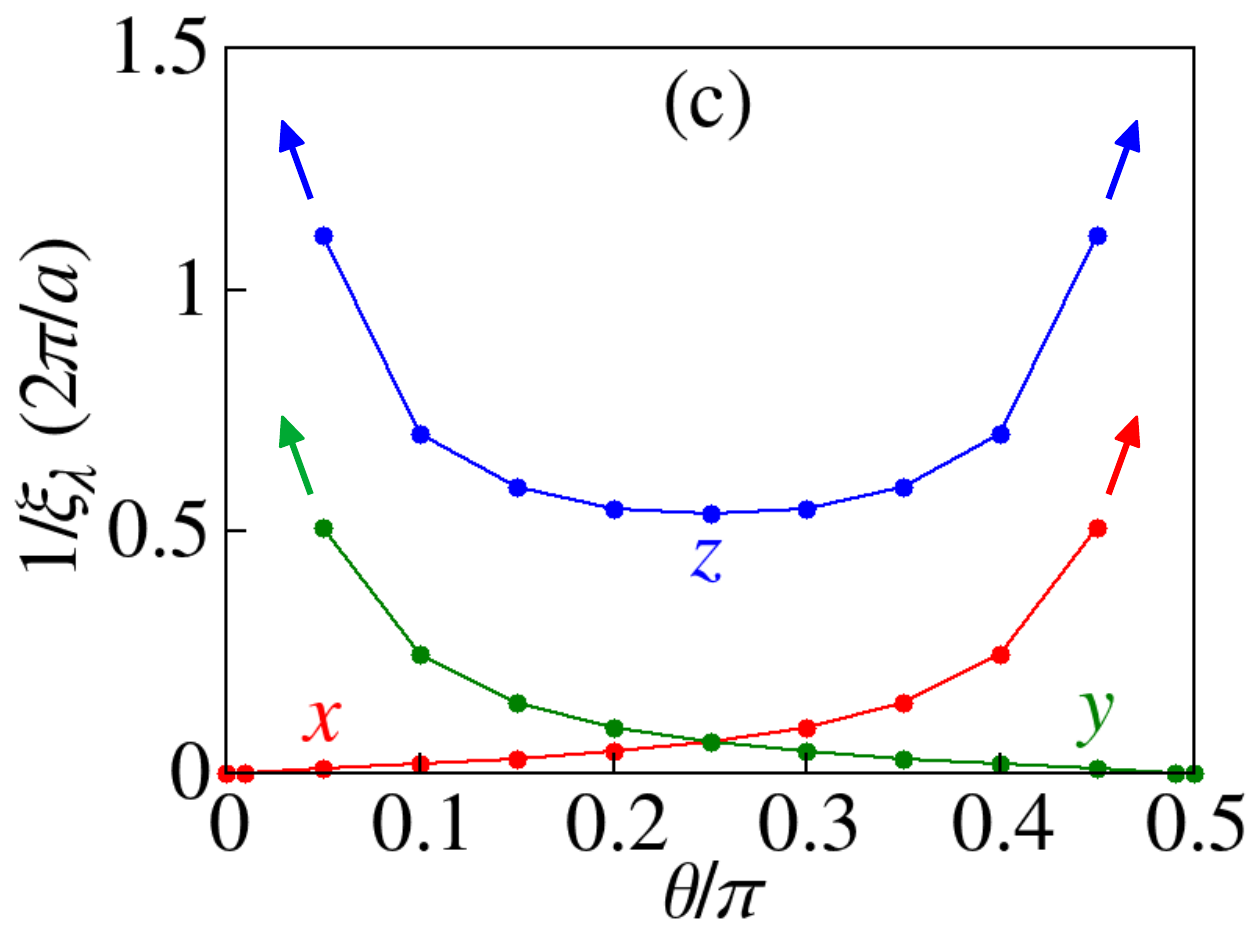}
\includegraphics[scale=0.2]{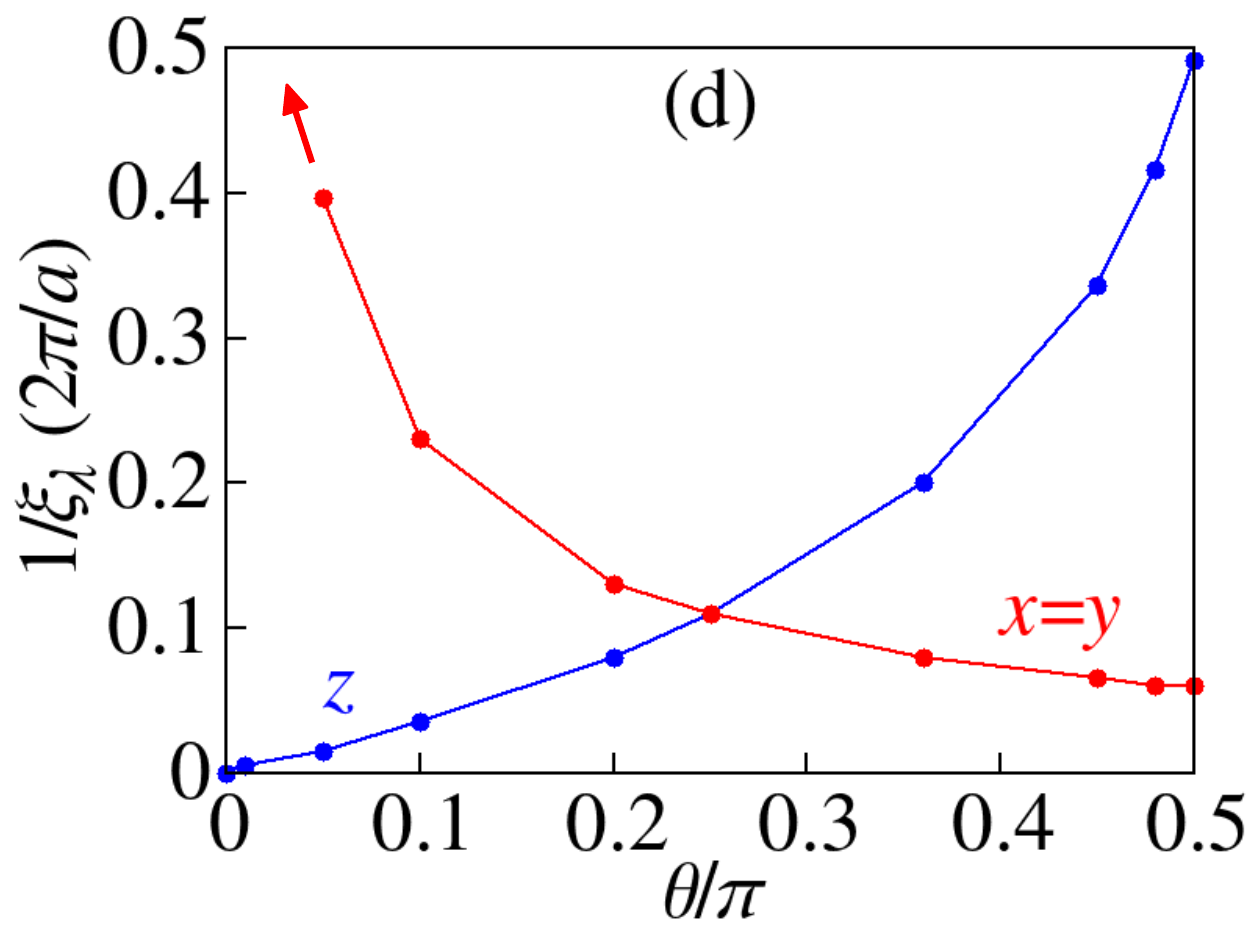}
\caption{\label{figure:width} (a) We calculate the half width at half maximum (HWHM) for each diagonal component of the static susceptibility along the cut $\mathbf{k}=(0,0,l)$ through the pinch point (see the inset where the cut is indicated by the white line). The example shown here is $\chi^{xx} (\mathbf{k})$ of the XY model \eqref{xymodel} at $\theta/\pi=0.25$, normalized by the maximum at $\chi^{xx} (\mathbf{k}=2\mathrm{X})$. (b) HWHM$_\lambda$ extracted from $\chi^{\lambda \lambda} (\mathbf{k})$ for $\lambda=x,y,z$ as a function of $\theta$ for the XY model \eqref{xymodel}. (c) The inverse correlation length $1/\xi_\lambda$, calculated from HWHM$_\lambda$ by subtracting a constant background \eqref{inverselength}, as a function of $\theta$ for the XY model \eqref{xymodel}. (d) The inverse correlation length $1/\xi_\lambda$ as a function of $\theta$ for the XXZ model \eqref{xxzmodel}. The arrows in (b-d) indicate divergences of HWHM$_{\mu , \nu}$ and $1/\xi_{\mu , \nu}$ in the Ising limit $(J_\lambda, J_\mu, J_\nu) = (1,0,0)$, where $\chi^{\mu \mu, \nu \nu} (\mathbf{k})$ are completely flat.}
\end{figure}

Let $(\lambda, \mu, \nu)$ be a cyclic permutation of $(x,y,z)$. As discussed in the main text, the static susceptibility $\chi^{\lambda \lambda} (\mathbf{k})$ in the $\pi$-flux QSI exhibits sharp or diffuse bow-tie patterns, with narrow or broadened pinch points, depending on whether the relative magnitude of the respective coupling $J_\lambda$ is large or small. In the classical spin ice limit at $(J_\lambda, J_\mu, J_\nu)=(1,0,0)$, which we call the $\lambda$-CSI, if we measure the equal-time spin structure factor $\mathcal{S}^{\lambda \lambda} (\mathbf{k})$ at finite temperatures, the intensity near a pinch point is known to take a Lorentzian form $\sim 1/(k^2 + \xi_\lambda^{-2})$, where $\xi_\lambda$ is interpreted as the correlation length of the ice rule being satisfied in the $S^\lambda$ basis \cite{PhysRevB.71.014424,science.1177582}. In other words, $\xi_\lambda$ gives the characteristic distance between $\lambda$-monopoles created by thermal fluctuations. Although we are using the static susceptibility to study quantum spin liquids at $T\longrightarrow 0$ here, it is sensible to look nonetheless at the width of the pinch point as a proximate measure of the inverse correlation length and infer the typical separation between excitations \cite{PhysRevX.9.011005}.

For concreteness, we study the XY and XXZ models
\begin{subequations}
\begin{align}
H_\mathrm{XY} &= \sum_{\langle ij \rangle} [ (\cos \theta) S_i^x S_j^x + (\sin \theta) S_i^y S_j^y ] , \label{xymodel} \\
H_\mathrm{XXZ} &= \sum_{\langle ij \rangle} [ (\cos \theta) S_i^z S_j^z + (\sin \theta) ( S_i^x S_j^x + S_i^y S_j^y ) ] , \label{xxzmodel}
\end{align}
\end{subequations}
with $\theta \in [0,\pi/2]$. For each $\lambda=x,y,z$, we calculate $\chi^{\lambda \lambda} (\mathbf{k}, \Lambda=\Lambda_f)$ along a one-dimensional cut $\mathbf{k}=(0,0,l)$ through $\mathbf{k}=2\mathrm{X}=(0,0,2)$, at which the pinch point is centered. To approximate the corresponding correlation length $\xi_\lambda$, we first extract the half width at half maximum (HWHM$_\lambda$) of the intensity along this cut \cite{PhysRevX.9.011005}, and plot it as a function of $\theta$, see Figs.~\ref{figure:width}a and \ref{figure:width}b. We note that the HWHM$_\lambda$ remains finite even in the $\lambda$-CSI limit, e.g., at $\theta=0$ and $\pi/2$ of $H_\mathrm{XY}$, where we ought to have $1/\xi_\lambda \longrightarrow 0$ as $T \longrightarrow 0$. We believe that this is an artefact due to correlation cutoffs and other approximations used in PFFRG. To correct for it, we define the inverse correlation length $1/\xi_\lambda$ by subtracting a constant background from HWHM$_\lambda$,
\begin{equation} \label{inverselength}
\xi_\lambda^{-1} = \mathrm{HWHM}_\lambda - \xi_0^{-1} ,
\end{equation}
with $1/\xi_0$ equal to the HWHM$_\lambda$ of $\lambda$-CSI. \eqref{inverselength} can also be understood as the statement that we are only interested in the change of HWHM$_\lambda$ relative to that of $\lambda$-CSI as we move away from the Ising limit. We remark that the truncation range of $L=6$ nearest neighbor bonds in our PFFRG calculations is greater than $\xi_0$, so subtracting the inverse of the former is not enough to yield a zero $1/\xi_\lambda$ for $\lambda$-CSI. Other approximations in PFFRG, e.g.~neglecting higher order vertices, might contribute to the background intensity as well.

The inverse correlation length calculated by \eqref{inverselength} is plotted as a function of $\theta$ for the models \eqref{xymodel} and \eqref{xxzmodel} in Figs.~\ref{figure:width}c and \ref{figure:width}d. These data suggest that we can interpolate smoothly from $x$-QSI to $y$-QSI or $z$-QSI via an XY-type model or through the Heisenberg point, while keeping the density of monopoles associated with the dominant interaction small (less than one monopole every four cubic unit cells). Indeed, tracking the lowest of $1/\xi_\lambda$, we see that it only grows to approximately $0.1 \times 2 \pi / a$ (at the Heisenberg point), which corresponds to a correlation length $\xi_\lambda \approx 1.59 a$ three to four times greater than the tetrahedral center-to-center distance $\sqrt{3}a/4 \approx 0.43 a$. The picture that emerges from this analysis is consistent with the scenario where the quantum spin liquids near the Ising limit, at the XX point, and at the Heisenberg point \footnote{We note however that several recent studies \cite{PhysRevB.105.054426,PhysRevLett.126.117204,PhysRevX.11.041021} point to a nematic order as the ground state of the pyrochlore Heisenberg antiferromagnet. Further work is needed to definitely settle this intriguing possibility.} are continuously connected in an extended $\pi$-flux QSI phase.

\bibliography{reference230901}

\end{document}